\documentclass[12pt]{iopart}
\usepackage{graphicx}
\usepackage{xcolor}

\begin{document}
\pdfminorversion=4 

\title[Asymmetry probe of the $Z\gamma$ process]{NNLO QCD predictions of the asymmetry probe of the $Z\gamma$ pair-production process}

\author{Kadir Saygin}

\address{Faculty of Aviation and Space Sciences, Necmettin Erbakan University, Konya, Turkey}
\ead{kadir.saygin@erbakan.edu.tr}
\vspace{10pt}
\begin{indented}
\item[]August 2023
\end{indented}

\begin{abstract}
The paper presents for the first time a novel idea of exploiting asymmetry between differential cross sections of the $Z\gamma$ pair-production in proton-proton ($pp$) collisions for the final states of a charged-lepton pair plus a photon $pp \rightarrow Z\gamma \rightarrow l^{+}l^{-}\gamma$ (leptonic decay) and of a neutrino pair plus a photon $pp \rightarrow Z \gamma \rightarrow \nu \bar{\nu}\gamma$ (invisible decay). Asymmetry between the leptonic and invisible decays of the $Z\gamma$ process is investigated by using fixed-order predictions through inclusion of next-to-next-to-leading (NNLO) radiative corrections in quantum chromodynamics (QCD) perturbation theory. NNLO QCD predictions are presented at various $pp$-collision energies as functions of several key observables including transverse momenta and azimuthal-angle separation, regarding the $Z\gamma$ decay products. The predicted distributions for the $Z\gamma$ asymmetry are provided based on realistic fiducial phase-space requirements in line with the related hadron-collider measurements. The predicted distributions are assessed at various $pp$-collision energies and in different phase-space regions such as with higher lepton-pair invariant mass $m^{l^{+}l^{-}}$ or higher neutrino-pair transverse momentum $p_{\rm{T}}^{\nu \bar{\nu}}$ requirements. The $Z\gamma$ asymmetry is shown to be significantly sensitive in different regions of phase space including high-$m^{l^{+}l^{-}}$ and high-$p_{\rm{T}}^{\nu \bar{\nu}}$ regions. The $Z\gamma$ asymmetry can therefore be translated into an important quantity for probing deviation from the Standard Model (SM) predictions. In this regard, the asymmetry probe is proposed as a sensitive indicator for indirect searches for physics beyond the SM encompassing high-mass resonances and dark-matter sector. 

\end{abstract}

%Uncomment for keywords
\vspace{2pc}

\textcolor{red}{\textbf{This is the version of the article before peer review or editing, as submitted by the author to the Physica Scripta. IOP Publishing Ltd is not responsible for any errors or omissions in this version of the manuscript or any version derived from it. The Version of Record is available online at https://doi.org/10.1088/1402-4896/ad1860}}

\maketitle

\section{Introduction}
\label{int}
\begin{figure}
\center
\includegraphics[width=6.3cm]{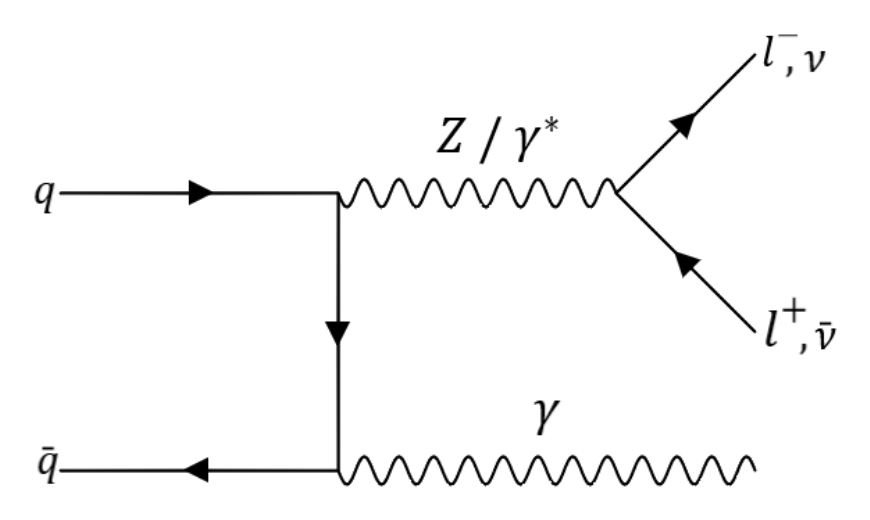}
\includegraphics[width=7.0cm]{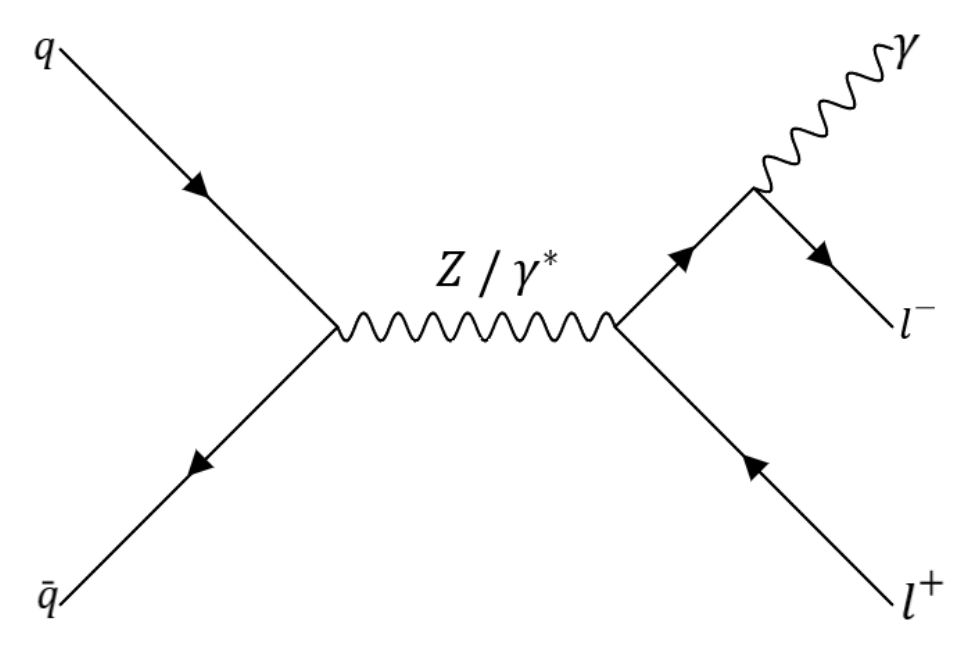}
\includegraphics[width=7.6cm]{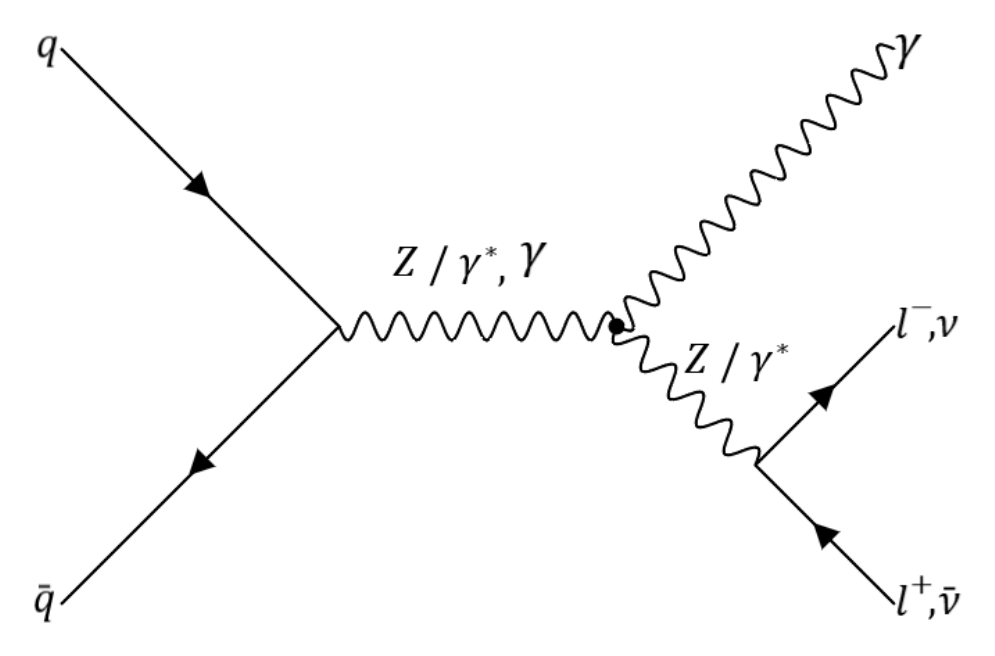}
\caption{Leading-order Feynman diagrams for the $Z\gamma$ process with initial-state photon radiation in both the leptonic and invisible decays (top-left), with final-state photon radiation in the leptonic decay (top-right), and for an aTGC vertex either $ZZ\gamma$ or $Z\gamma\gamma$ (bottom) which is forbidden in the SM at tree level.}
\label{fig:1}      
\end{figure}

The $Z\gamma$ pair-production at hadron colliders represents a prominent process for tests of the electroweak (EW) sector of the Standard Model (SM) and for searches for new-physics effects. In proton-proton ($pp$) collisions, the $Z\gamma$ process primarily proceeds through production of a $Z$ boson in association with a photon which is radiated off either initial-state quarks or final-state decay products, depending on particular $Z$-boson decay channel. The $Z$ boson can decay hadronically into a quark pair $Z\rightarrow q\bar{q}$, leptonically into a charged-lepton pair $Z\rightarrow l^{+}l^{-}$\footnote[1]{Throughout the paper, the $Z$ boson stands for the $Z / \gamma^{*}$ regarding charged-lepton pair decay, where it also corresponds to contribution from the $\gamma^{*} \rightarrow l^{+}l^{-}$ process and its interference with the leptonic $Z$-boson decay.}, and invisibly into a neutrino pair $Z\rightarrow \nu \bar{\nu}$, where hadronic decay is less preferential by experiments owing to higher background contamination over leptonic and invisible decays. In the SM, direct coupling of $Z$ bosons to photons is forbidden at tree level, but can be featured in theories beyond the SM that predict gauge-boson self-interactions. The $Z\gamma$ process is mainly used for probing anomalous (nonstandard) trilinear gauge-boson couplings (aTGCs) $ZZ\gamma$ and $Z\gamma\gamma$, which can in turn be translated into evidence for new-physics effects. Sample leading-order Feynman diagrams of the $Z\gamma$ process for the leptonic $Z\gamma \rightarrow l^{+}l^{-}\gamma$ and invisible $Z\gamma \rightarrow \nu \bar{\nu}\gamma$ decays, and additionally for an aTGC vertex are shown in Figure~\ref{fig:1}. The $ZZ\gamma$ and $Z\gamma\gamma$ vertices are described theoretically by using both the vertex-function approach based on four coupling parameters $h_{i}^{V}$ with $i=1,...,4$ and $V\in\{Z,\gamma\}$ and the Lagrangian approach in extensions of the SM gauge structure~\cite{Baur:1992cd,DeFlorian:2000sg,Gounaris:1999kf}. Both approaches are equivalent and are included directly into the framework of the effective-field theory (EFT)~\cite{Degrande:2012wf,Degrande:2013kka}. Measurements of production cross sections of the $Z\gamma$ process were reported by several experiments from $e^{+}e^{-}$ collisions at the CERN Large-Electron Positron (LEP) collider~\cite{OPAL:2000ijr,L3:2004hlr,DELPHI:2007gzg} and from $p\bar{p}$ collisions at the Fermilab Tevatron collider~\cite{D0:2009olv,CDF:2011rqn,D0:2011tjg}. Measurements of the $Z\gamma$ process in both leptonic and invisible decay modes are reported from $pp$ collisions at the CERN Large Hadron Collider (LHC) at center-of mass energies of $\sqrt{s}=$7 TeV and 8 TeV by the ATLAS and CMS Collaborations~\cite{ATLAS:2013way,CMS:2013yrt,CMS:2013ryd,CMS:2015wtk,CMS:2016cbq,ATLAS:2016qjc}, as well as more recently at 13 TeV by the ATLAS Collaboration~\cite{ATLAS:2018nci,ATLAS:2019gey}. All these measurements have been observed to be in agreement with the SM predictions, revealing no evidence for the presence of aTGCs $ZZ\gamma$ and $Z\gamma\gamma$, whereby the LHC experiments set the most restrictive limits on such couplings to date.    

The $Z\gamma$ process of the SM is further important in various respects in searches for new-physics models. New gauge bosons can manifest themselves as new resonances in kinematic distributions which can be probed by exploiting the $Z\gamma$ decays. The ATLAS and CMS Collaborations at the LHC have carried out several searches for new resonances and presented stringent constraints on various new-physics models~\cite{ATLAS:2014lfk,CMS:2016ssv,CMS:2016wgx,ATLAS:2016mti,CMS:2017dyb}, where non-resonant $Z\gamma$ decays stand for the dominant background process in these searches. In addition, the $Z\gamma$ process constitutes major background in searches for the Higgs-boson decay with a very small branching fraction $H\rightarrow Z \gamma$ at the LHC~\cite{CMS:2013rmy,ATLAS:2014fxe,ATLAS:2017zd,CMS:2018myz,ATLAS:2020qcv,CMS:2022ahq}. The $H\rightarrow Z \gamma$ process is not only important for extraction of Higgs-boson properties but also for searches for new-physics models including supersymmetry and extended Higgs sectors, as elaborated such as in the Refs.~\cite{Abbasabadi:1996ze,Chiang:2012qz,Chen:2013vi,Hammad:2015eca,Dedes:2019bew}. Moreover, the invisible decay of the $Z\gamma$ process is an irreducible background to dark-matter searches in the photon plus missing transverse energy final state~\cite{ATLAS:2020uiq}. Such invisible final-state signatures might indicate dark-matter candidates produced in pair as $\chi \bar{\chi}$ in view of the interpretations of simplified models of dark-matter searches at the LHC~\cite{Abdallah:2015ter}. In this regards, experimental studies of the $Z\gamma$ process involving both the SM and beyond-SM measurements benefit directly from high-precision and high-accuracy computations of production cross sections in the perturbation theory. As demanded by the LHC experiments, computations of this process in underlying perturbation theories have advanced significantly in the recent years to include EW corrections, available through next-to-leading order (NLO) accuracy~\cite{Hollik:2004tm,Accomando:2005ra,Denner:2015fca}, and to standardize inclusion of quantum chromodynamics (QCD) corrections at next-to-NLO (NNLO) accuracy as such in the more recent phenomenological studies~\cite{Grazzini:2013bna,Grazzini:2015nwa,Campbell:2017aul}. More improved theoretical predictions of particular kinematics of the $Z\gamma$ process are also achieved by means of the matched computations of NNLO QCD corrections and resummed or parton-shower calculations~\cite{Wiesemann:2020gbm,Lombardi:2020wju,Lombardi:2021wug}.           

This paper presents for the first time asymmetry between differential cross sections of the leptonic and invisible decays of the $Z\gamma$ process in $pp$ collisions, $pp \rightarrow Z\gamma \rightarrow l^{+}l^{-}\gamma$ and $pp \rightarrow Z\gamma \rightarrow \nu \bar{\nu}\gamma$, respectively, by using genuine theoretical predictions through inclusion of NNLO radiative corrections in QCD perturbation theory. Differential cross-section predictions for the $Z\gamma$ process as functions of a number of important observables are first compared with the relevant LHC measurements at 13 TeV. Differential distributions of cross-section ratio and asymmetry of the $Z\gamma$ process are predicted as functions a set of key observables including transverse momenta and azimuthal-angle separation of the $Z\gamma$ decay products, and compared at various $pp$-collision energies of 13 TeV, 14 TeV, and 100 TeV. $Z\gamma$ asymmetry distributions are predicted using fiducial phase-space regions with realistic experimental selection requirements as well as with higher lepton-pair invariant mass or higher neutrino-pair transverse momentum requirements. Sensitivity of $Z\gamma$ asymmetry distributions is assessed extensively in different regions of phase space, where these distributions can be used for probing any deviation from the known SM domain which might manifest signs for new-physics models. $Z\gamma$ asymmetry probe is consequently proposed as a novel idea in comparison to the existing studies of searches for aTGCs in the $Z\gamma$ process, which can represent sensitive indicator in indirect searches for physics beyond the reach of the SM involving high-mass resonances and dark-matter candidates. 

\section{Methodology}
\label{meth}
\subsection{Computational preliminaries}
\label{computational}
Various aspects of the computational setup which is exploited in the course of this paper can be summarized in the followings. Fully differential cross-section calculations of the off-shell $Z\gamma$ production in the leptonic $l^{+}l^{-}\gamma$ and invisible $\nu \bar{\nu}\gamma$ decay modes in $pp$ collisions are performed using the computational framework MATRIX (version 2.1.0.beta2)~\cite{Grazzini:2013bna,Grazzini:2015nwa,Grazzini:2017mhc}. Fixed-order perturbative QCD calculations are achieved using the Catani-Seymour dipole-subtraction method at NLO~\cite{Catani:1996jh,Catani:1996vz} and using the \emph{$q_{\rm{T}}$}-subtraction method~\cite{Catani:2007vq,Catani:2012qa} at NNLO within the MATRIX framework. A finite cut-off value for the slicing parameter $r_{\rm{cut}}$ of the \emph{$q_{\rm{T}}$}-subtraction method is employed as $r_{\rm{cut}}=0.0015 (0.15\%)$ to regularize soft and collinear divergences of the real radiation in the perturbation expansion. The NNLO QCD computations require evaluation of tree-level scattering and one-loop amplitudes and of one-loop squared and two-loop corrections to the Born subprocess $q\bar{q} \rightarrow l^{+}l^{-}\gamma$, which are provided by the OpenLoops 2 generator as incorporated into the MATRIX framework~\cite{Gehrmann:2011ab,Cascioli:2011va,Denner:2016kdg,Buccioni:2017yxi,Buccioni:2019sur}. In addition to the usual $q\bar{q}$ channels, contribution from the loop-induced gluon fusion channel to the $Z\gamma$ process at leading order ($gg \rightarrow Z\gamma$) is included in the NNLO computations. This contribution is checked that it has an impact amounting to 1.5\% on integrated fiducial cross sections of both the leptonic and invisible decays. Parton distribution functions (PDFs) of the proton are included in the computations relying on the NNPDF4.0 modeling~\cite{NNPDF:2021njg}. The NNPDF40\_(n)nlo\_as\_01180  PDF sets are used at appropriate perturbative order (N)NLO in which the strong coupling constant is set to its standard value $\alpha_{S}(m_{Z})=$ 0.118. The computations entail evaluation of the PDF sets which is handled with the LHAPDF (v6.4.0) platform~\cite{Buckley:2014ana}. 

In the computations, the $Z$-boson mass $m_{Z}$ is defined using the complex-mass scheme~\cite{Denner:2005fg}. The EW mixing angle $\sin^{2}\theta_{W}$ (and hence the electroweak coupling $\alpha$) are expressed in terms of complex $W$- and $Z$-boson mass terms in this scheme. Input parameters for the EW couplings are supplemented in the computations according to the so-called $G_{\mu}$ input scheme. The input values of EW parameters are used identically as the PDG values~\cite{ParticleDataGroup:2022pth} for the $m_{Z}$, the decay width $\Gamma_{Z}$, and the Fermi constant $G_{\rm{F}}$ as $m_{Z}$=91.1876 GeV, $\Gamma_{Z}$=2.4952 GeV, and $G_{\rm{F}}$= 1.16639 $\times10^{-5}$ GeV$^{-2}$, respectively. The renormalization $\mu_{R}$ and factorization $\mu_{F}$ scales are required to be set up in calculations of hadronic cross section of the $Z\gamma$ process. Central values $\mu_{0}$ of the $\mu_{R}$ and $\mu_{F}$ scales are set up dynamically to the invariant mass of the system of colorless final states, i.e., $\mu_{R}=\mu_{F}= \mu_{0}=m^{l^{+}l^{-}\gamma}$ in the leptonic decay and $\mu_{R}=\mu_{F}= \mu_{0}=m^{\nu \bar{\nu}\gamma}$ in the invisible decay. Transverse mass of the final-state products is checked as an alternate central value $\mu_{0}=m_{T}^{l^{+}l^{-}\gamma}=m_{T}^{\nu \bar{\nu}\gamma}$ for the scales, leading only to a difference less than subpercent level on integrated fiducial cross sections in comparison with the current invariant-mass choice. Furthermore, quarks and antiquarks (except for the top quark) and gluons in the initial state are treated massless based on a number of flavor scheme $N_{F}=$5. Additionally leptons except for tau are considered massless by a default option in the computational setup.       

\subsection{Fiducial phase-space requirements} 
\label{fiducial}
The baseline fiducial phase-space selection for the $Z\gamma$ process is chosen to be realistic in line with the ATLAS measurements of differential cross sections at 13 TeV in the leptonic~\cite{ATLAS:2019gey} and in the invisible ~\cite{ATLAS:2018nci} decay modes. In the leptonic decay, same-flavor leptons with opposite electric charge signs at Born-level (i.e., no lepton dressing with collinear photons is imposed on leptons) are required, where lepton indicates either an electron or a muon $l\in\{e,\mu\}$, not sum over them. Leading and subleading leptons are required to have transverse momenta $p_{\rm{T}}^{l_{1}}>$ 30 GeV and $p_{\rm{T}}^{l_{2}}>$ 25 GeV, respectively, in the lepton pseudorapidity acceptance $|\eta^{l}|<$ 2.47. Photon is required to have $p_{\rm{T}}^{\gamma}>$ 30 GeV in the acceptance $|\eta^{\gamma}|<$ 2.37. Photon and leptons must be separated within a cone of radius $\Delta R(l,\gamma)>$ 0.4\footnote[2]{Separation cuts between the particles $i$ and $j$ are implemented based on the definition $\Delta R(i,j)=\sqrt{\Delta \eta (i,j)^{2}+\Delta \phi (i,j)^{2}}$, where $\Delta \eta (i,j)$ and $\Delta \phi (i,j)$ stand for differences in pseudorapidity and azimuthal angle of these particles, respectively.}. Invariant mass of lepton-pair is required to be $m^{l^{+}l^{-}}>$ 40 GeV to avoid low-mass resonance effects. In addition, sum of invariant masses of lepton-pair and lepton-pair-photon systems is required to fulfill the requirement $m^{l^{+}l^{-}}+m^{l^{+}l^{-}\gamma}>$ 182 GeV for the purpose of suppressing low-$p_{\rm{T}}^{\gamma}$ photon emission from final-state leptons. In the invisible decay, missing transverse momentum $p_{\rm{T}}^{miss}$ corresponding to transverse momentum of neutrino pair is selected to be $p_{\rm{T}}^{\nu \bar{\nu}}>$ 150 GeV, where $\nu$ ($\bar{\nu}$) indicates each type of neutrino (antineutrino) $\nu \in\{\nu_{e},\nu_{\mu}\}$, not sum over them. No pseudorapidity requirement is imposed on neutrinos. Photon is required to have $p_{\rm{T}}^{\gamma}>$ 150 GeV in the acceptance $|\eta^{\gamma}|<$ 2.37. In both leptonic and invisible decays, emission of accompanying jets in final states is allowed, where jets are defined by using the anti-$k_t$ clustering algorithm~\cite{Cacciari:2008gp} with cone radius parameter $\Delta R =$ 0.4. No explicit requirement is imposed on jets in the leptonic decay, whereas jets are required to be inclusive with $p_{\rm{T}}^{j}>$ 50 GeV in the acceptance $|\eta^{j}|<$ 4.5 in the invisible decay. In the invisible decay, additionally a separation requirement between jets and photon is applied with a cone of radius requirement $\Delta R(j,\gamma)>$ 0.3.

In the $Z\gamma$ final-states, photons do not only originate from direct production mechanisms in the hard process, besides they are produced through fragmentation of a QCD parton. Contribution from fragmented photons to the $Z\gamma$ production cross section requires evaluation of a non-perturbative fragmentation function which yields typically large uncertainties. A photon isolation method in this regard is applied to suppress contribution from fragmented photons, which can be considered as part of the baseline fiducial selection. The Frixione smooth-cone isolation prescription~\cite{Frixione:1998jh} is implemented in the computations, which provides complete suppression for the fragmentation contribution. In this isolation prescription, total amount of transverse energy of partons $E_{T}$ inside a cone of radius $R$ centered around the photon direction is required to be less than a maximum value $E^{max}_{T}(R)$, where it is defined as $E^{max}_{T}(R) \equiv \epsilon_{\gamma} p_{\rm{T}}^{\gamma} \left[(1-\cos R)/(1-\cos R_{max})\right]^{n}$. The isolation criterion $E_{T}<E^{max}_{T}(R)$ is required for all cone sizes, provided that $R\leq R_{max}$. The isolation parameters are used identically from the ATLAS measurements~\cite{ATLAS:2019gey,ATLAS:2018nci} as $n=$ 2 (1), $\epsilon_{\gamma}=0.1$, and $R_{max}=0.1$ for the leptonic (invisible) decay. The baseline fiducial selection is only used for validation of differential cross-section predictions using the ATLAS results at 13 TeV. The baseline fiducial selection is simply referred to as the $\emph{Baseline selection--Set I}$ in the paper text and summarized in Table~\ref{tab:1}.

In addition to the $\emph{Baseline selection--Set I}$, fiducial phase-space requirements are slightly modified to have more robust and common kinematics for the final states of the $Z\gamma$ process. In the leptonic decay, leptons are required to have a symmetric threshold as $p_{\rm{T}}^{l_{1},l_{2}}>$ 30 GeV. Invariant mass requirement involving lepton-pair-photon systems $m^{l^{+}l^{-}}+m^{l^{+}l^{-}\gamma}>$ 182 GeV is removed to take into account additional photons from final-state leptons. In the invisible decay, selection on neutrino pair and photon is loosened for a wider phase-space region by the requirements $p_{\rm{T}}^{\nu \bar{\nu}}>$ 60 GeV and $p_{\rm{T}}^{\gamma}>$ 30 GeV, respectively. Even though final-state jets are allowed in the leptonic and invisible decays as in the $\emph{Baseline selection--Set I}$, explicit requirements on jets including the separation $\Delta R(j,\gamma)$ cut are removed in the invisible decay to bring fiducial selections of the both decays in line with each other regarding jet kinematics. The choices for the photon isolation parameters are applied exactly the same for the both decays. These modified-baseline fiducial phase-space requirements are referred to as the $\emph{Baseline selection--Set II}$ in the text for brevity and summarized in Table~\ref{tab:2}. The $\emph{Baseline selection--Set II}$ is used to obtain the main NNLO QCD predictions of the paper.             

\begin{table*}
\renewcommand{\arraystretch}{1.5}
\caption{\label{tab:1}Summary of the baseline fiducial phase-space requirements, the $\emph{Baseline selection--Set I}$, which is used in validation of the NNLO QCD predictions with the measurements} 
\lineup
\resizebox{15.6cm}{!}{%
\begin{tabular}{@{}*{11}{c|cc|cc}}
\br
criteria & \multicolumn{2}{c|}{$Z\gamma \rightarrow l^{+}l^{-}\gamma$} & \multicolumn{2}{c}{$Z \gamma \rightarrow \nu \bar{\nu}\gamma$}  \\
\mr
lepton cuts & \multicolumn{2}{c|}{$p_{\rm{T}}^{l_{1}}$$>$30 GeV, $p_{\rm{T}}^{l_{2}}$$>$25 GeV, $|\eta^{l}|$$<$2.47} & \multicolumn{2}{c}{---}  \\
neutrino cuts & \multicolumn{2}{c|}{---} & \multicolumn{2}{c}{$p_{\rm{T}}^{\nu \bar{\nu}}$$>$150 GeV}  \\
photon cuts & \multicolumn{2}{c|}{$p_{\rm{T}}^{\gamma}$$>$30 GeV, $|\eta^{\gamma}|$$<$2.37} & \multicolumn{2}{c}{$p_{\rm{T}}^{\gamma}$$>$150 GeV, $|\eta^{\gamma}|$$<$2.37}  \\
jet cuts & \multicolumn{2}{c|}{---} & \multicolumn{2}{c}{$p_{\rm{T}}^{j}$$>$50 GeV, $|\eta^{j}|$$<$4.5}  \\
invariant-mass cuts & \multicolumn{2}{c|}{$m^{l^{+}l^{-}}$\!$>$40 GeV, $m^{l^{+}l^{-}}$\!+$m^{l^{+}l^{-}\gamma}$$>$182 GeV} & \multicolumn{2}{c}{---}  \\
separation cuts & \multicolumn{2}{c|}{$\Delta R(l,\gamma)$$>$0.4} & \multicolumn{2}{c}{$\Delta R(j,\gamma)$$>$0.3}  \\
\mr
smooth-cone photon isolation & \multicolumn{2}{c|}{$n$=2, $\epsilon_{\gamma}$=0.1, $R_{max}$=0.1} & \multicolumn{2}{c}{$n$=1, $\epsilon_{\gamma}$=0.1, $R_{max}$=0.1}  \\
\br
\end{tabular}%
}
\end{table*}

\begin{table*}
\renewcommand{\arraystretch}{1.5}
\caption{\label{tab:2}Summary of the modified-baseline fiducial phase-space requirements, the $\emph{Baseline selection--Set II}$, which is used to obtain the main NNLO QCD predictions} 
\lineup
\resizebox{15.6cm}{!}{%
\begin{tabular}{@{}*{11}{c|cc|cc}}
\br
criteria & \multicolumn{2}{c|}{$Z\gamma \rightarrow l^{+}l^{-}\gamma$} & \multicolumn{2}{c}{$Z \gamma \rightarrow \nu \bar{\nu}\gamma$}  \\
\mr
lepton cuts & \multicolumn{2}{c|}{$p_{\rm{T}}^{l_{1}}$$>$30 GeV, $p_{\rm{T}}^{l_{2}}$$>$30 GeV, $|\eta^{l}|$$<$2.4} & \multicolumn{2}{c}{---}  \\
neutrino cuts & \multicolumn{2}{c|}{---} & \multicolumn{2}{c}{$p_{\rm{T}}^{\nu \bar{\nu}}$$>$60 GeV}  \\
photon cuts & \multicolumn{2}{c|}{$p_{\rm{T}}^{\gamma}$$>$30 GeV, $|\eta^{\gamma}|$$<$2.4} & \multicolumn{2}{c}{$p_{\rm{T}}^{\gamma}$$>$30 GeV, $|\eta^{\gamma}|$$<$2.4}  \\
jet cuts & \multicolumn{2}{c|}{---} & \multicolumn{2}{c}{---}  \\
invariant-mass cuts & \multicolumn{2}{c|}{$m^{l^{+}l^{-}}$\!$>$40 GeV} & \multicolumn{2}{c}{---}  \\
separation cuts & \multicolumn{2}{c|}{$\Delta R(l,\gamma)$$>$0.4} & \multicolumn{2}{c}{---}  \\
\mr
smooth-cone photon isolation & \multicolumn{2}{c|}{$n$=1, $\epsilon_{\gamma}$=0.1, $R_{max}$=0.1} & \multicolumn{2}{c}{$n$=1, $\epsilon_{\gamma}$=0.1, $R_{max}$=0.1}  \\
\br
\end{tabular}%
}
\end{table*}

\subsection{Definition of the $Z\gamma$ asymmetry}
\label{asymmetry}
In the computations, cross-section calculations at (N)NLO accuracy are carried out differentially as functions of important observables for both the leptonic and invisible decays. The ratio of differential cross sections of these decays $R_{Z\gamma}$ is an important quantity in controlling neutrino pairs, which leave a detector volume undetected resulting in significantly large missing transverse energy, relative to lepton pairs as being well-reconstructed experimental objects. The $R_{Z\gamma}$ can additionally be used to probe any deviation of branching fractions from the SM predictions, such as at higher collision energies or in unexplored phase-space regions. Further in this regard, an asymmetry quantity $A_{Z\gamma}$ can be constructed, exploiting differential cross sections $\textrm{d}\sigma$ of these decays with respect to each other, regarding the $Z \gamma$ process. The asymmetry $A_{Z\gamma}$ is expressed in terms of the $R_{Z\gamma}$, differential in an observable $\lambda$ as 

\begin{equation}
\label{eqn:1}   
A_{Z\gamma}=\frac{R_{Z\gamma}-1}{R_{Z\gamma}+1},\: \textrm{where} \: R_{Z\gamma}=\frac{d\sigma(Z\gamma \rightarrow \nu \bar{\nu}\gamma)/d\lambda}{d\sigma(Z\gamma \rightarrow l^{+}l^{-}\gamma)/d\lambda} .
\end{equation}
In general the $A_{Z\gamma}$ helps to correlate the leptonic and invisible decays via differential cross sections relative to each other in the $Z\gamma$ process. The $A_{Z\gamma}$ can specifically be used to improve controlling of the $Z\gamma$ process, for instance when it constitutes a major background in direct new-physics searches. It can also be benefited to probe any small departure from the SM in indirect searches which might indicate signs for new phenomena. The $A_{Z\gamma}$ is constructed as a double-ratio of the differential cross sections, meaning that it can act as a more sensitive probe for the leptonic and invisible decays of the $Z\gamma$ process in comparison to the $R_{Z\gamma}$. In addition, the $A_{Z\gamma}$ can allow for cancellation of residual systematic uncertainties in experimental studies that are more correlated in it its double-ratio structure over the $R_{Z\gamma}$. As a consequence, the $A_{Z\gamma}$ turns into an important quantity to enable stringent tests in high-precision experimental and theoretical studies. The asymmetry probe $A_{Z\gamma}$ has analogous definition with the definition of the lepton charge asymmetry of the $W$-boson production such as in Refs.~\cite{Ocalan:2021htz,Ocalan:2022zbn}. Besides, it is related with the asymmetry definition as a probe for the invisible $Z$-boson process in the absence of a final-state photon as published earlier in the Ref.~\cite{Saygin:2023xae}.                

In calculations of the $R_{Z\gamma}$ and the $A_{Z\gamma}$, theoretical uncertainties due to missing higher-order QCD corrections beyond NNLO are estimated using the 7-point variations scheme of the QCD scales $\mu_{R}$ and $\mu_{F}$. In this scheme, differential cross sections with $\mu_{R}$ and $\mu_{F}$ are varied up and down around the central scale $\mu_{0}$ with the following factors

\begin{equation}
\label{eqn:2}   
(\mu_{R}, \mu_{F}) \in \{(2,2), (2,1), (1,2), (1,1), (1,1/2), (1/2,1), (1/2,1/2)\} \cdot  \mu_{0}.
\end{equation}
Largest uncertainty from these variations is taken as theoretical scale uncertainty which is properly propagated to differential distributions of the $R_{Z\gamma}$ and $A_{Z\gamma}$. In the numerator and denominator of the $R_{Z\gamma}$ and of the $A_{Z\gamma}$, scale uncertainties are considered uncorrelated in a conservative approach and quoted by taking into account standard deviations. Besides that uncertainties due to PDF modeling and $\alpha_{S}$ variation are not included in the predictions, as computations of these uncertainties are not automated in the MATRIX. These uncertainties are therefore treated fully correlated in the ratios of the $R_{Z\gamma}$ and the $A_{Z\gamma}$.    

\section{Phenomenological results}
\label{pheno}

\subsection{Comparison with the recent data}
\label{comp}
Differential cross-section distributions of the $Z\gamma$ process are predicted at (N)NLO accuracy in perturbative QCD for validation purposes. Differential distributions are predicted for the leptonic $pp \rightarrow Z\gamma \rightarrow l^{+}l^{-}\gamma$ and the invisible $pp \rightarrow Z\gamma \rightarrow \nu \bar{\nu}\gamma$ decay modes at 13 TeV by using fiducial phase-space requirements of the $\emph{Baseline selection--Set I}$ as summarized in Table~\ref{tab:1}. The predicted distributions are compared with the recent ATLAS measurements at 13 TeV~\cite{ATLAS:2018nci,ATLAS:2019gey} for both the decay modes. Theoretical uncertainties due to variations in the QCD scales are included in the predictions, while total experimental uncertainties including both statistical and systematic uncertainties are quoted for the data distributions. The predicted distributions are provided as functions of important kinematical observables using the same choices of bin edges from the reference ATLAS measurements to enable direct comparisons. In the prediction-to-data comparisons, a lepton corresponds to either an $e$ or a $\mu$, whereas a neutrino is exceptionally treated to refer to sum over all neutrino flavors apart from its treatment in the rest of the paper.  

\begin{figure}
\center
\includegraphics[width=10.0cm]{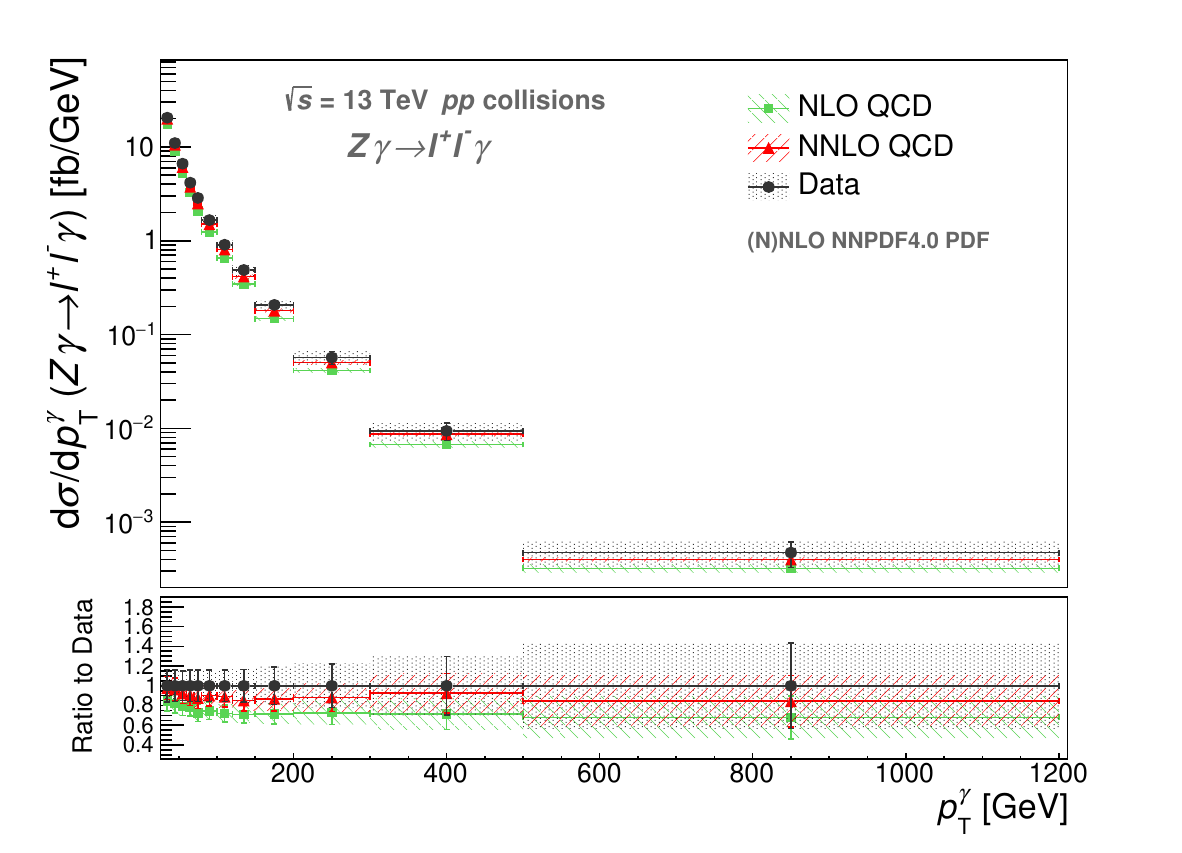}
\includegraphics[width=10.0cm]{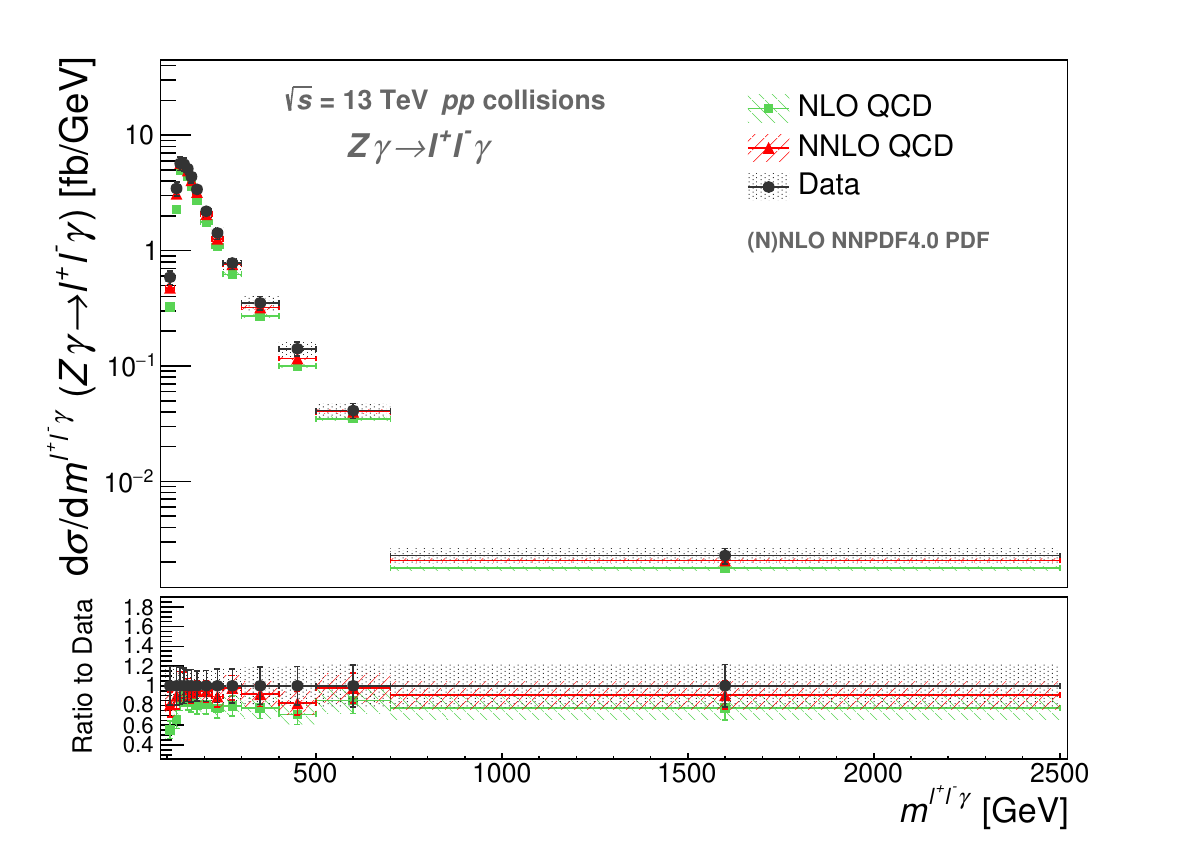}
\caption{The predicted differential cross-section distributions at (N)NLO accuracy in comparison with the ATLAS data~\cite{ATLAS:2019gey} at 13 TeV. The predictions are compared with the data for the key observables $p_{\rm{T}}^{\gamma}$ (top) and $m^{l^{+}l^{-}\gamma}$ (bottom) of the leptonic decay of the $Z\gamma$ process $Z\gamma \rightarrow l^{+}l^{-}\gamma$. Theoretical uncertainties due to scale variations are quoted for the predictions, while total experimental uncertainties are quoted for the data distributions. The ratios of the predictions to the data are shown in the lower inset.}
\label{fig:2}      
\end{figure}

The predicted distributions are first compared with the data in the leptonic decay for the key observables involving the transverse momentum of the photon $p_{\rm{T}}^{\gamma}$ and the invariant mass of the lepton-pair-photon system $m^{l^{+}l^{-}\gamma}$ in the ranges 30--1200 GeV and 95--2500 GeV, respectively, as shown in Figure~\ref{fig:2}. The predictions are acquired at parton level without accounting for soft QCD effects from hadronization and underlying events, and thus parton-to-particle level correction factors are applied to the predictions as multiplicative factors to bring them in line with the data measurement at particle level. These correction factors are adopted from the ATLAS measurement~\cite{ATLAS:2019gey}, which vary in the ranges $0.93\pm0.006$--$0.97\pm0.004$ for the $p_{\rm{T}}^{\gamma}$ and $0.92\pm0.005$--$0.95\pm0.002$ for the $m^{l^{+}l^{-}\gamma}$ distributions. The prediction distributions are observed to be in better agreement with the data with the inclusion of NNLO corrections in comparison with the predictions at NLO. Precision in the predictions of the differential cross sections is improved in going from NLO to NNLO as anticipated. Therefore, NNLO QCD predictions are justified to be reliable in description of the data distributions within the quoted uncertainties for the leptonic decay of the $Z\gamma$ process. 

\begin{figure}
\center
\includegraphics[width=10.0cm]{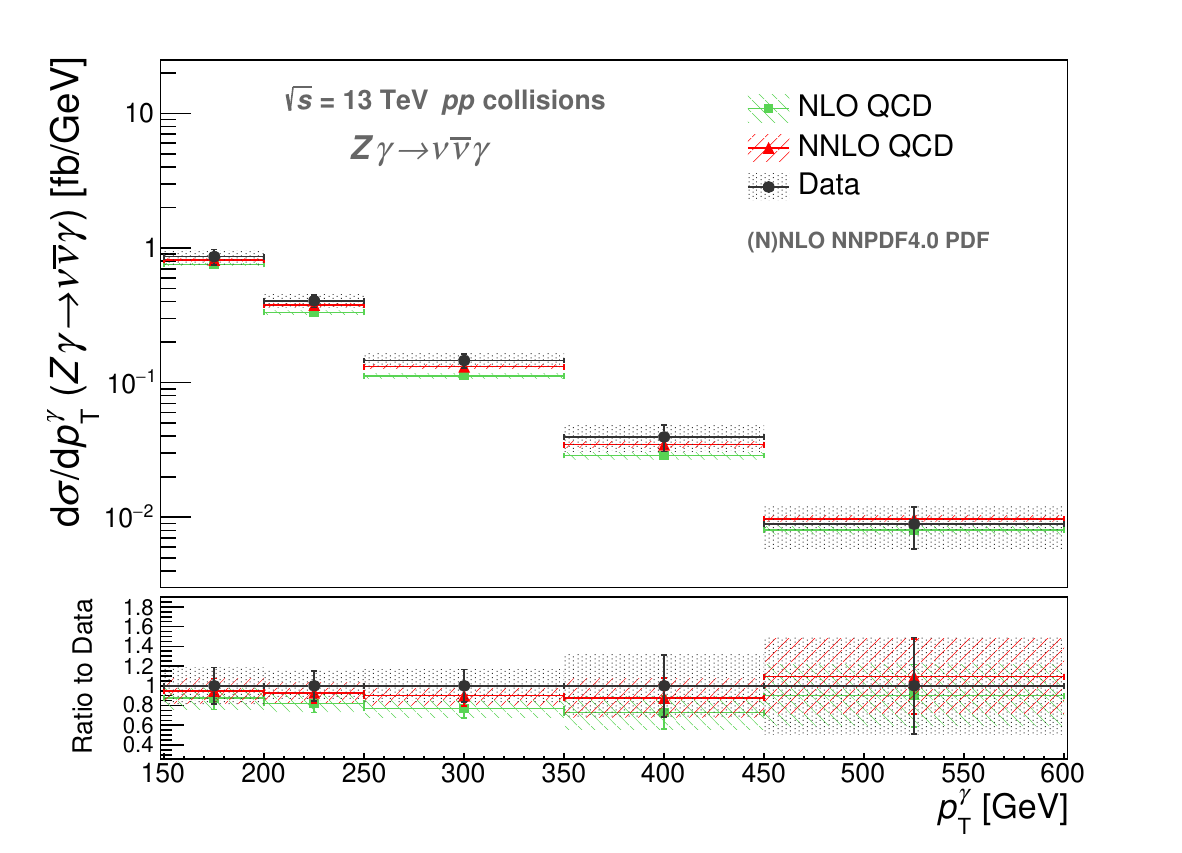}
\includegraphics[width=10.0cm]{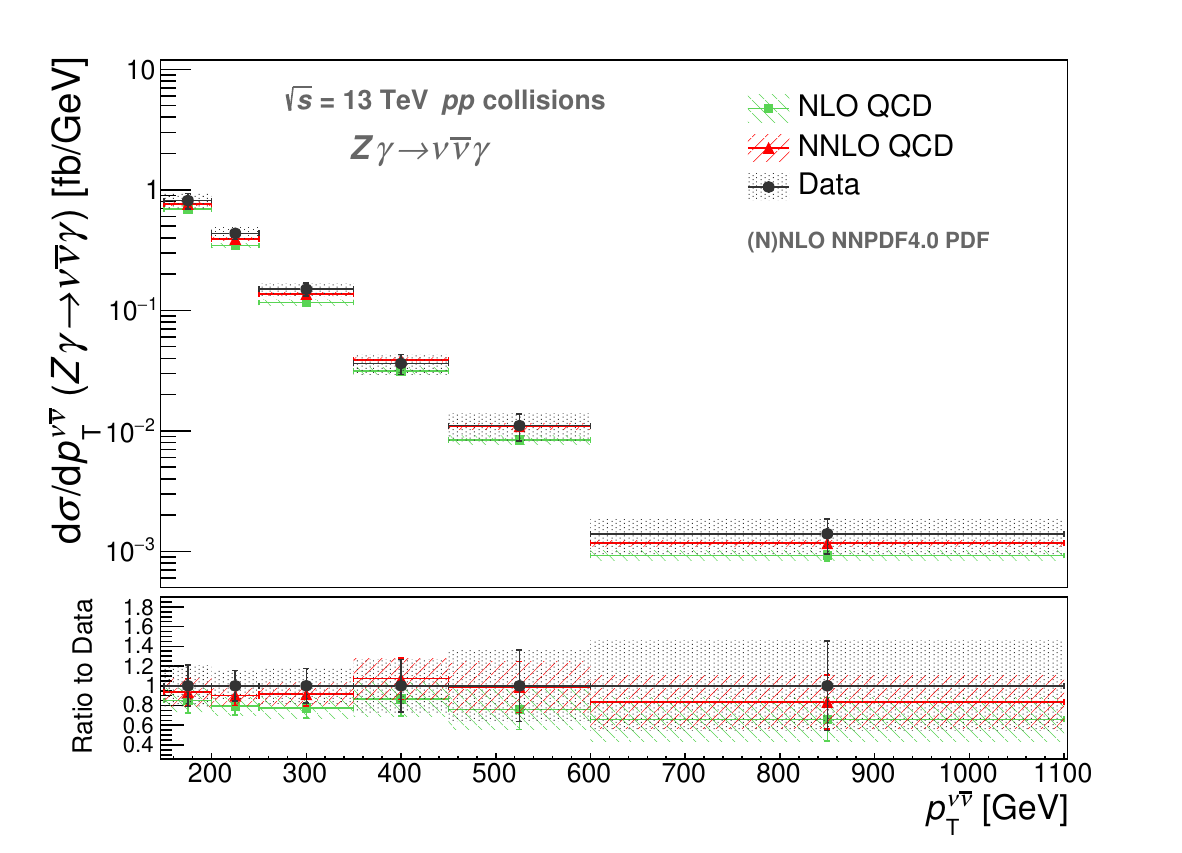}
\caption{The predicted differential cross-section distributions at (N)NLO accuracy in comparison with the ATLAS data~\cite{ATLAS:2018nci} at 13 TeV. The predictions are compared with the data for the key observables $p_{\rm{T}}^{\gamma}$ (top) and $p_{\rm{T}}^{\nu \bar{\nu}}$ (bottom) of the invisible decay of the $Z\gamma$ process $Z\gamma \rightarrow \nu \bar{\nu}\gamma$. Theoretical uncertainties due to scale variations are quoted for the predictions, while total experimental uncertainties are quoted for the data distributions. The ratios of the predictions to the data are shown in the lower inset.}
\label{fig:3}      
\end{figure}

Similarly the predicted distributions at (N)NLO are compared with the data in the invisible decay for the key observables involving the $p_{\rm{T}}^{\gamma}$ and the transverse momentum of the neutrino pair $p_{\rm{T}}^{\nu \bar{\nu}}$ in the ranges 150--600 GeV and 150--1100 GeV, respectively. The bin range 600--1100 GeV for the $p_{\rm{T}}^{\gamma}$ is omitted from the comparison owing to very large statistical uncertainty in the data distribution. The predictions at parton level are corrected with the parton-to-particle level correction factor $0.87\pm0.004$, which is provided by the ATLAS measurement~\cite{ATLAS:2018nci}, and compared with the particle-level data distributions on an equal basis. The predictions are provided in Figure~\ref{fig:3} in comparison with the data for the invisible final state accompanied by inclusive jets, such that $N_{jets}\geq$ 0. Description of the data distributions is significantly improved in going from NLO to NNLO accuracy in the predictions. Consequently, NNLO QCD predictions are justified as providing fairly good description of the data distributions within the uncertainties, as well as for the invisible decay of the $Z\gamma$ process.    

\subsection{The $R_{Z\gamma}$ and $A_{Z\gamma}$ predictions at collider energies}
\label{predictions}
Differential distributions of the $R_{Z\gamma}$ and $A_{Z\gamma}$ as functions of a set of important observables are predicted at NNLO accuracy in QCD perturbation theory. The predicted distributions are acquired at various $pp$-collision energies of the LHC and of the future hadron colliders. The High-Luminosity LHC (HL-LHC)~\cite{ZurbanoFernandez:2020cco} and the Future Circular Collider devoted to a hadron collider (FCC-hh)~\cite{FCC:2018vvp} are the two future $pp$ colliders that are conceived to accumulate large amount of data at 14 TeV and 100 TeV energies, respectively. Sensitivities to higher collision energies in addition to 13 TeV are of importance to unveil future physics potential of the $Z\gamma$ process which can be enhanced by the $R_{Z\gamma}$ and in particular by the $A_{Z\gamma}$. In comparisons of the predicted distributions at different energies, fiducial phase-space requirements of the $\emph{Baseline selection--Set II}$, as summarized in Table~\ref{tab:2}, are consistently applied for the leptonic and invisible decay modes. The predictions are provided as functions of the $p_{\rm{T}}^{Z}$, $p_{\rm{T}}^{\gamma}$, $p_{\rm{T}}^{Z\gamma}$, and $\Delta \phi (Z,\gamma)$ of the $Z\gamma$ decay products. The $p_{\rm{T}}^{Z}$ ($p_{\rm{T}}^{Z\gamma}$) is equivalent of transverse momenta of lepton-pair (lepton-pair-photon) system for the leptonic decay and of neutrino-pair (neutrino-pair-photon) system of the invisible decay. The $\Delta \phi (Z,\gamma)$ corresponds to the azimuthal-angle separation between lepton-pair (neutrino-pair) system and photon in the respective leptonic (invisible) decay. The observables are defined at the parton level and accordingly no parton-to-particle level correction is applied. Theoretical scale uncertainties are reported in the predicted distributions of the observables.  

\begin{figure}
\center
\includegraphics[width=10.0cm]{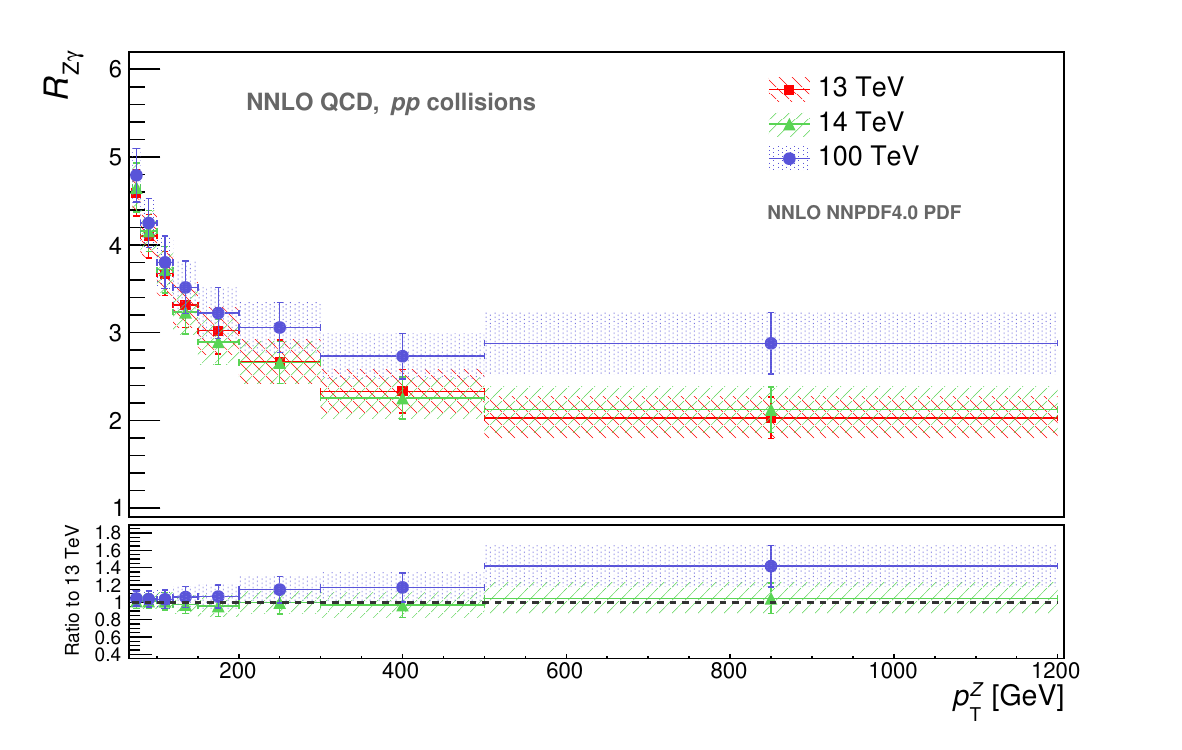}
\includegraphics[width=10.0cm]{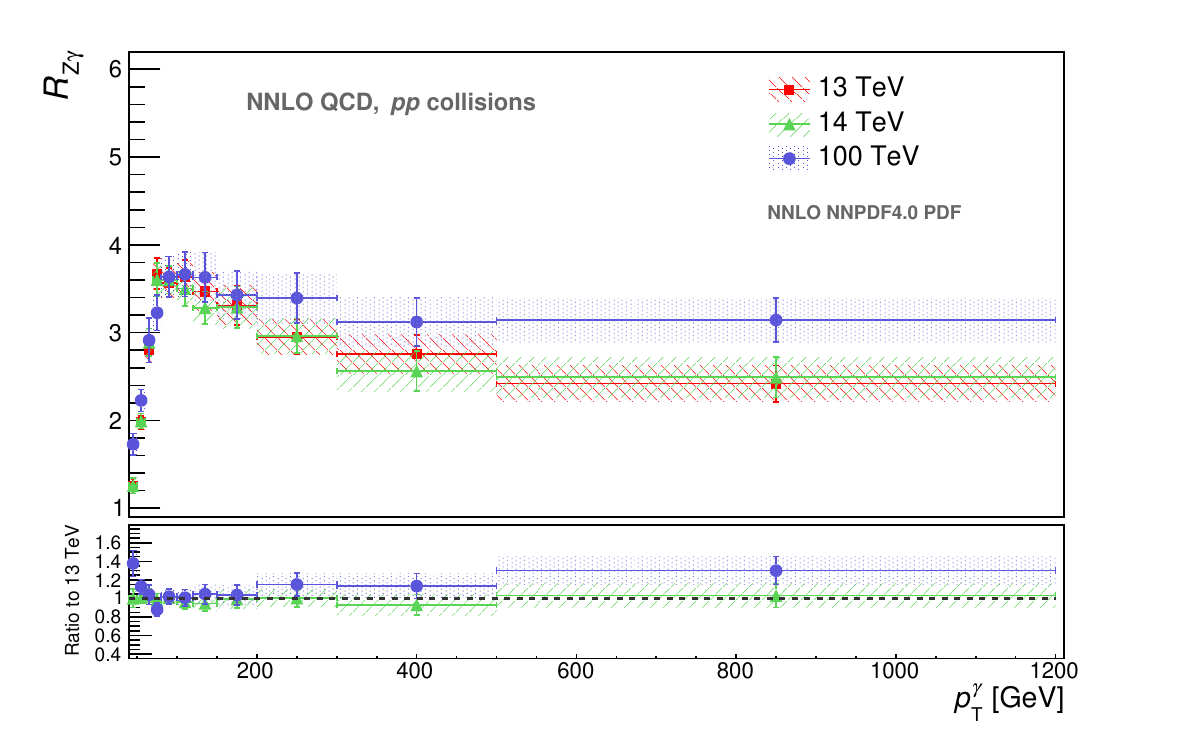}
\caption{The $R_{Z\gamma}$ distributions for the $p_{\rm{T}}^{Z}$ (top) and the $p_{\rm{T}}^{\gamma}$ (bottom) predicted at NNLO QCD accuracy at 13 TeV (LHC), 14 TeV (HL-LHC), and 100 TeV (FCC-hh) $pp$ collision energies. Theoretical uncertainties due to scale variations are included for the predictions. The ratios of the predictions at higher energies to the predictions at 13 TeV are provided in the lower inset.}
\label{fig:4}      
\end{figure}

\begin{figure}
\center
\includegraphics[width=10.0cm]{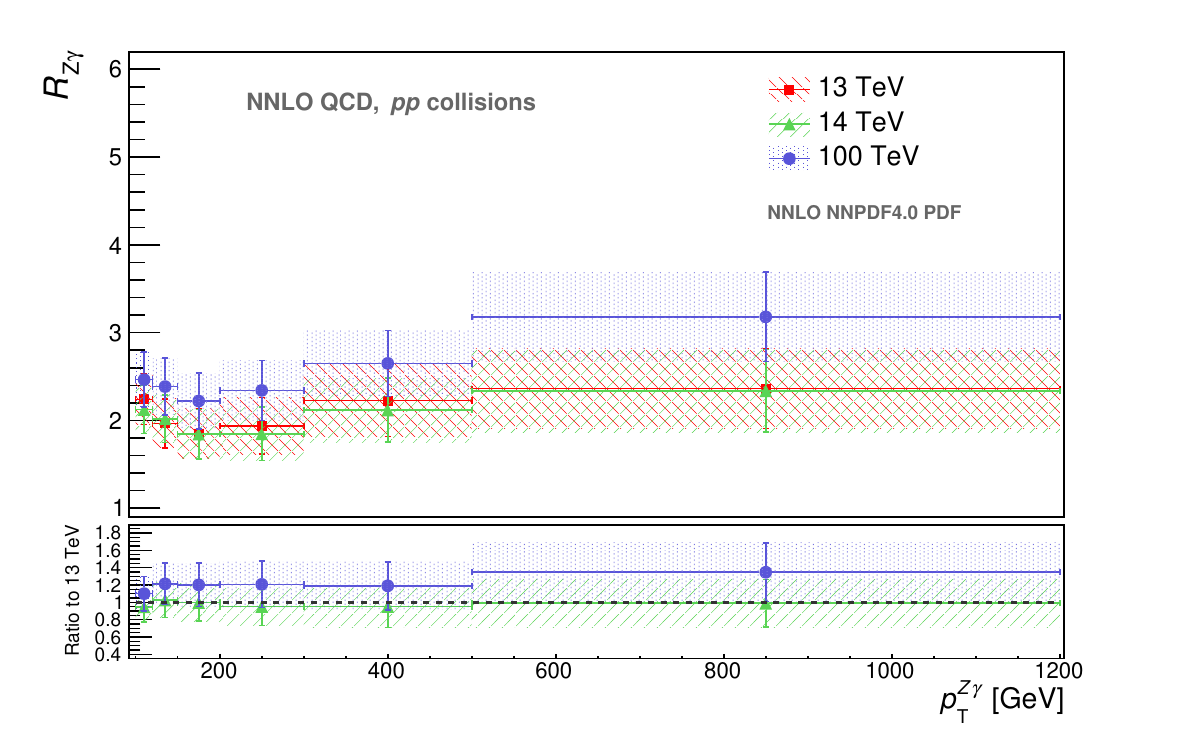}
\includegraphics[width=10.0cm]{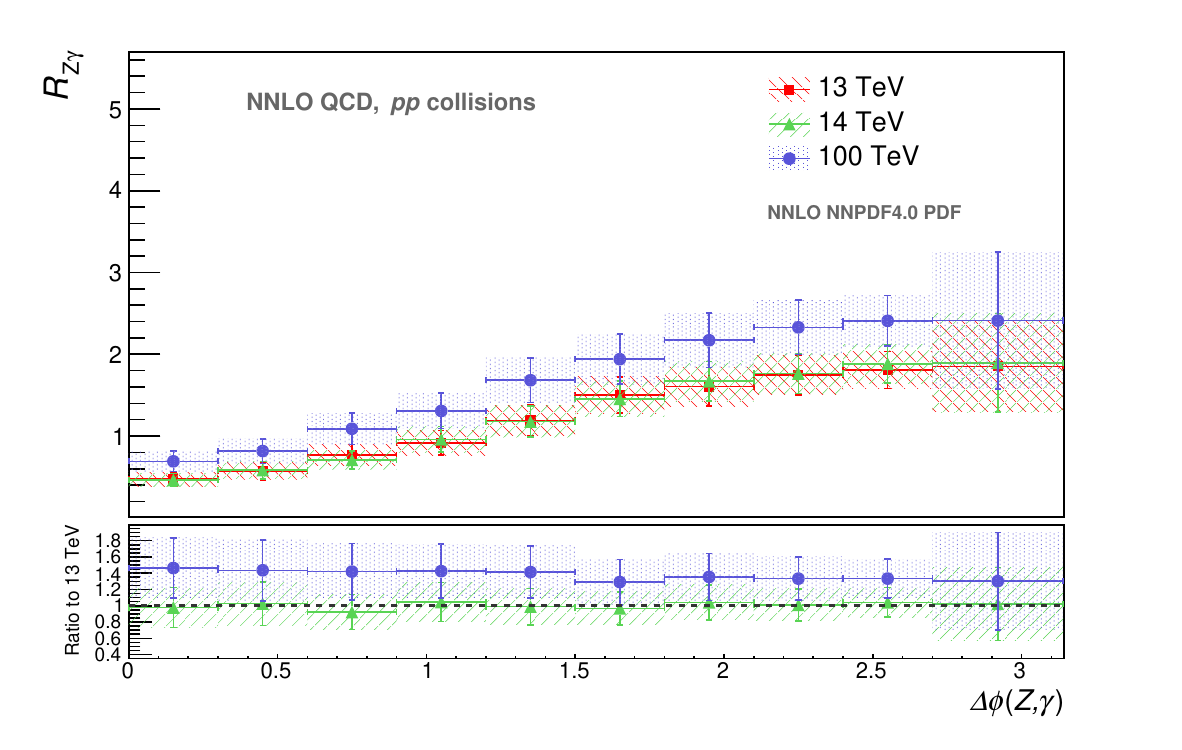}
\caption{The $R_{Z\gamma}$ distributions for the $p_{\rm{T}}^{Z\gamma}$ (top) and the $\Delta \phi (Z,\gamma)$ (bottom) predicted at NNLO QCD accuracy at 13 TeV (LHC), 14 TeV (HL-LHC), and 100 TeV (FCC-hh) $pp$ collision energies. Theoretical uncertainties due to scale variations are included for the predictions. The ratios of the predictions at higher energies to the predictions at 13 TeV are provided in the lower inset.}
\label{fig:5}      
\end{figure}

The predicted differential distributions at 13 TeV, 14 TeV, and 100 TeV are first compared for the $R_{Z\gamma}$ as functions of the key observables $p_{\rm{T}}^{Z}$, $p_{\rm{T}}^{\gamma}$, $p_{\rm{T}}^{Z\gamma}$, and $\Delta \phi (Z,\gamma)$ as presented in Figure~\ref{fig:4} and Figure~\ref{fig:5}. Difference between the $R_{Z\gamma}$ predictions at 13 TeV and 14 TeV is not much pronounced, whereas difference among the predictions at lower energies and the predictions at 100 TeV is substantial for wide ranges of the observables. The $R_{Z\gamma}$ distributions for all the observables have increasing trend from lower energies towards 100 TeV. The $R_{Z\gamma}$ is observed to increase more towards higher ranges of the kinematical observables, while it increases almost at constant rates in the entire range of the $\Delta \phi (Z,\gamma)$, in going from lower energies to 100 TeV. The $R_{Z\gamma}$ distributions tend to increase more significantly at 100 TeV in the phase-space region enclosed by the conditions $p_{\rm{T}}^{Z}>$ 200 GeV and $p_{\rm{T}}^{\gamma}>$ 200 GeV. In general the $R_{Z\gamma}$ increases substantially at 100 TeV leading to considerable enhancement in the invisible decay which would render the $R_{Z\gamma}$ a sensitive quantity for searches for weakly-interacting particles as dark-matter candidates in the future.         

\begin{figure}
\center
\includegraphics[width=10.0cm]{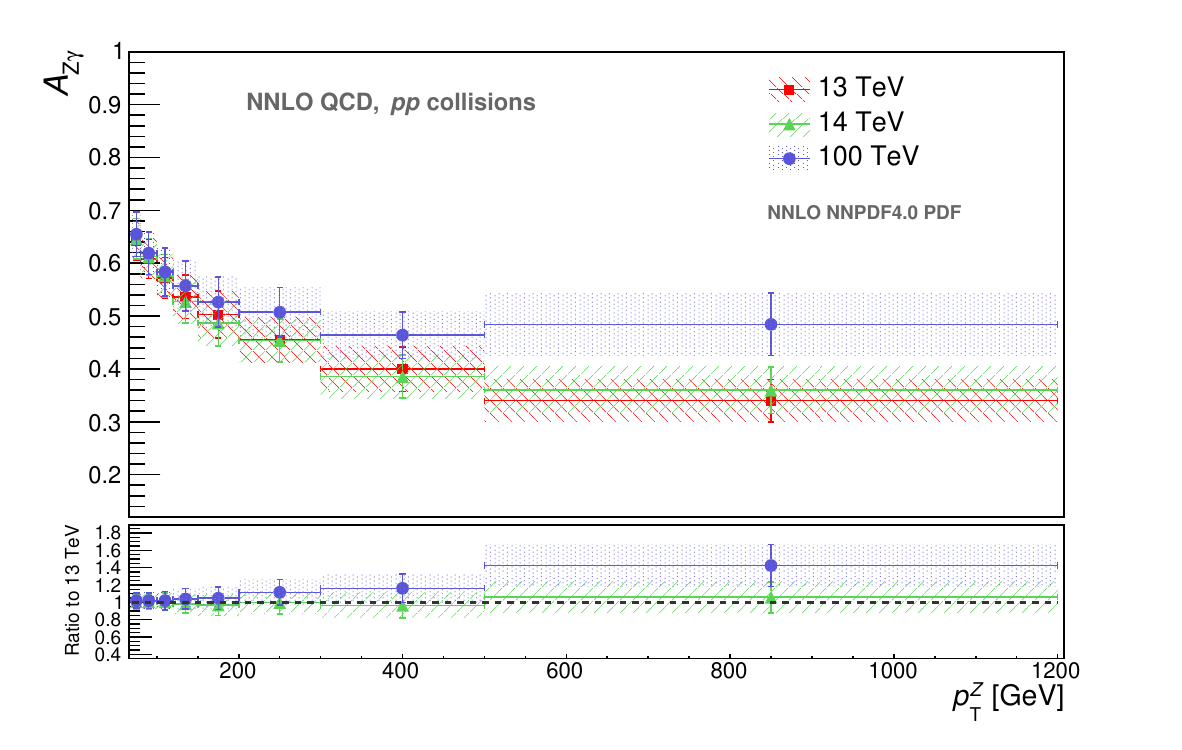}
\includegraphics[width=10.0cm]{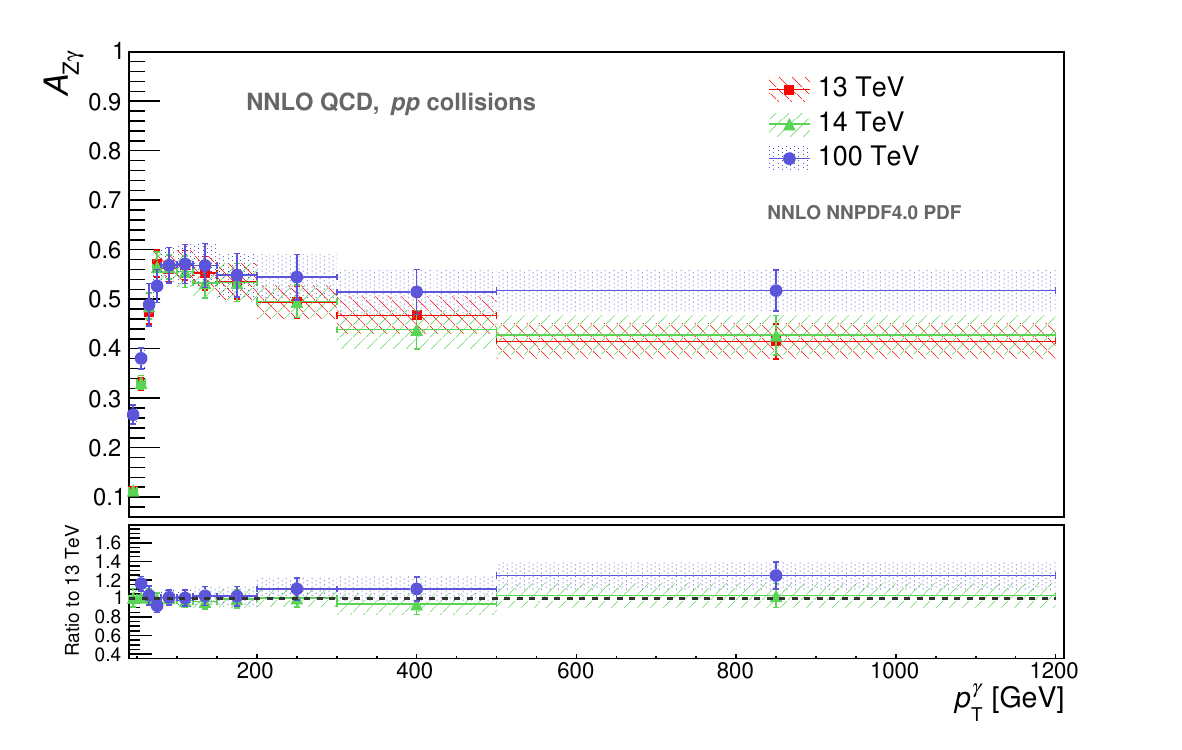}
\caption{The $A_{Z\gamma}$ distributions for the $p_{\rm{T}}^{Z}$ (top) and the $p_{\rm{T}}^{\gamma}$ (bottom) predicted at NNLO QCD accuracy at 13 TeV (LHC), 14 TeV (HL-LHC), and 100 TeV (FCC-hh) $pp$ collision energies. Theoretical uncertainties due to scale variations are included for the predictions. The ratios of the predictions at higher energies to the predictions at 13 TeV are provided in the lower inset.}
\label{fig:6}      
\end{figure}

\begin{figure}
\center
\includegraphics[width=10.0cm]{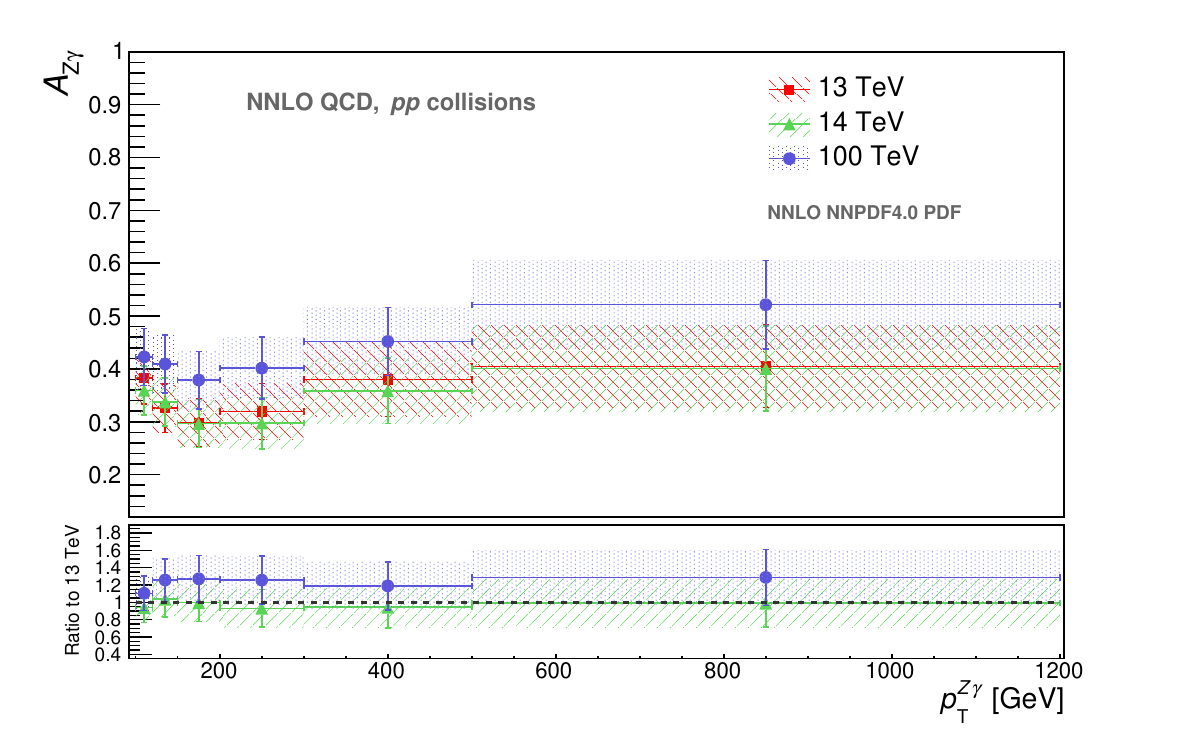}
\includegraphics[width=10.0cm]{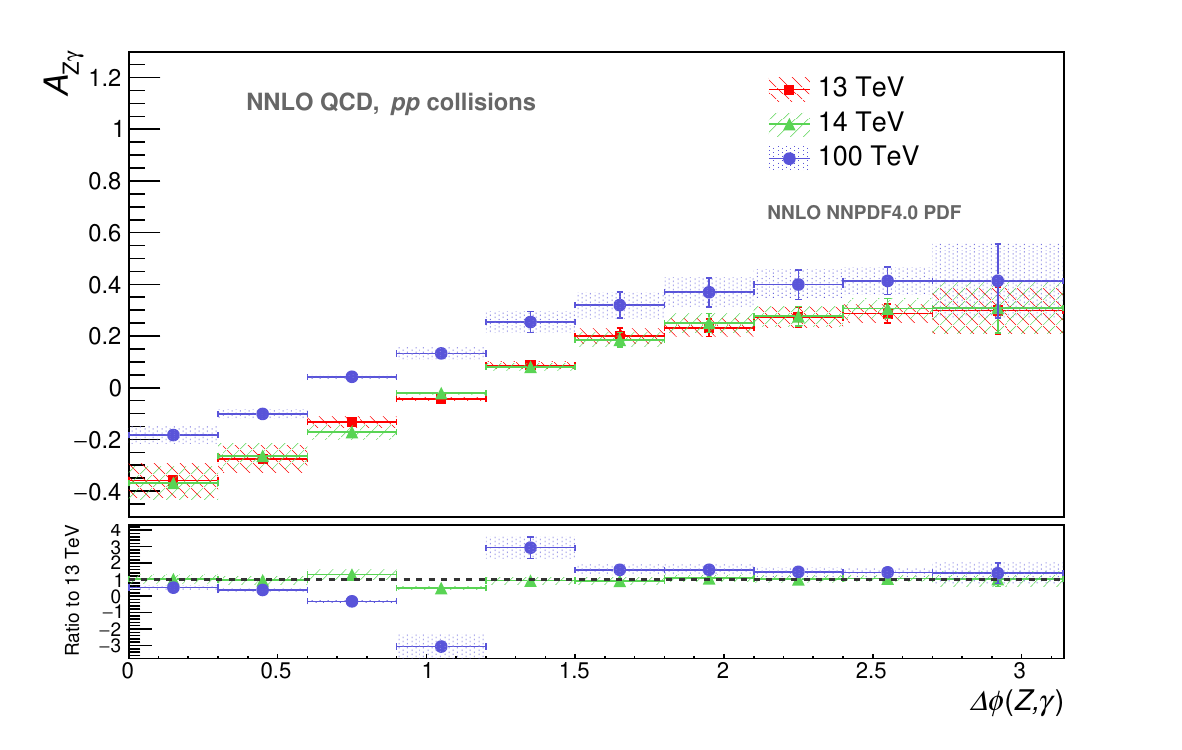}
\caption{The $A_{Z\gamma}$ distributions for the $p_{\rm{T}}^{Z\gamma}$ (top) and the $\Delta \phi (Z,\gamma)$ (bottom) predicted at NNLO QCD accuracy at 13 TeV (LHC), 14 TeV (HL-LHC), and 100 TeV (FCC-hh) $pp$ collision energies. Theoretical uncertainties due to scale variations are included for the predictions. The ratios of the predictions at higher energies to the predictions at 13 TeV are provided in the lower inset.}
\label{fig:7}      
\end{figure}

Next the predicted distributions of the $A_{Z\gamma}$ as functions of the $p_{\rm{T}}^{Z}$, $p_{\rm{T}}^{\gamma}$, $p_{\rm{T}}^{Z\gamma}$, and $\Delta \phi (Z,\gamma)$ are compared at 13 TeV, 14 TeV, and 100 TeV in Figure~\ref{fig:6} and Figure~\ref{fig:7}. The $A_{Z\gamma}$ distributions are predicted to be comparable at 13 TeV and 14 TeV regarding most of the ranges of the observables, while they are observed to increase considerably in going from lower energies towards 100 TeV for all the observables. Moreover, the distributions tend to be more flat and higher in the phase-space region specified by the conditions $p_{\rm{T}}^{Z}>$ 200 GeV and $p_{\rm{T}}^{\gamma}>$ 200 GeV at 100 TeV in contrast to the distributions at 13 TeV and 14 TeV. The $A_{Z\gamma}$ is predicted to be higher for the complete range of the $p_{\rm{T}}^{Z\gamma}$ at 100 TeV in comparison with the predictions at lower energies. Furthermore, the $A_{Z\gamma}$ is predicted to be negative for the region $\Delta \phi (Z,\gamma)<$ 1.2 (0.6) at 13 TeV and 14 TeV (100 TeV). It is observed to have consistently higher distribution shape for the $\Delta \phi (Z,\gamma)$ at 100 TeV with respect to the shapes at lower energies, nevertheless its distribution ratio at 100 TeV to 13 TeV fluctuates in the range $\Delta \phi (Z,\gamma)=$ 0.6--1.5 and becomes flat and higher for the higher ranges $\Delta \phi (Z,\gamma)>$ 1.5. By these token, the $A_{Z\gamma}$ distributions are assessed to be sensitive to higher-collision energies towards 100 TeV for all the observables. Sensitivities of the $R_{Z\gamma}$ and the $A_{Z\gamma}$ to higher-collision energies are not exactly at the same level but are generally assessed to be comparable for the kinematical observables. Notably, the $A_{Z\gamma}$ exhibits different sensitivity to increasing collision energy with the $\Delta \phi (Z,\gamma)$ in comparison to the $R_{Z\gamma}$, where more stringent tests can be performed with the $A_{Z\gamma}$ using this angular observable. Consequently the $A_{Z\gamma}$ can be used as a good probe for checking any experimental deviation from the predictions at various collider energies including those in the future, which would mean a sensitive indicator to signs for new-physics models including dark matter.     

\subsection{The $A_{Z\gamma}$ probe for high-mass region}
\label{high-mass}
The $A_{Z\gamma}$ predictions are of interest in more constrained phase-space region of the $Z\gamma$ process with higher $m^{l^{+}l^{-}}$ thresholds for searches for high-mass resonances beyond the SM. In this respect the $A_{Z\gamma}$ is predicted at NNLO using the $\emph{Baseline selection--Set II}$ from Table~\ref{tab:2}, taking into consideration various $m^{l^{+}l^{-}}$ requirements in the leptonic decay mode as $m^{l^{+}l^{-}}>$ 40 GeV, 100 GeV and 500 GeV. Sensitivity of the $A_{Z\gamma}$ to higher $m^{l^{+}l^{-}}$ requirements is assessed by comparing the predictions at 13 TeV. The predictions are obtained at parton level and are presented with theoretical scale uncertainties. First of all, the $A_{Z\gamma}$ is predicted in bins of the invariant mass of the $Z\gamma$ process $m^{Z \gamma}$, which is equivalent for the invariant mass of the lepton-pair-photon system $m^{l^{+}l^{-}\gamma}$ of the leptonic decay and for the invariant mass of the neutrino-pair-photon system $m^{\nu \bar{\nu}\gamma}$ of the invisible decay. The predicted numerical values in bins of the $m^{Z \gamma}$ up to 2500 GeV are provided in Table~\ref{tab:3}. The $A_{Z\gamma}$ is negative in the low-$m^{Z \gamma}$ bins 95--150 GeV and then increase the most up to 0.52 in the bin 400-500 GeV in the $m^{l^{+}l^{-}}>$ 40 GeV region. The $A_{Z\gamma}$ is positive in the $m^{l^{+}l^{-}}>$ 100 GeV region with the values between $\sim$0.72--0.99, decreasing towards the highest $m^{Z \gamma}$ range up to 2500 GeV. In the $m^{l^{+}l^{-}}>$ 500 GeV region, the numerical values are available in the $m^{Z \gamma}$ range 500--2500 GeV, which also follow a decreasing trend towards the highest bin. In going from the $m^{l^{+}l^{-}}>$ 40 GeV region to the highest region $m^{l^{+}l^{-}}>$ 500 GeV, the $A_{Z\gamma}$ clearly follows an increasing pattern based on available numerical values depending on the $m^{l^{+}l^{-}}$ thresholds. In high-mass resonance searches the predicted numerical values can be considered as a probe for comparison with experimental measurements. Any excess over the predictions in bins of the $m^{Z \gamma}$ might indicate new-physics effects including any non-SM boson which decays into the $Z\gamma$ pair with a narrow-width invariant-mass distribution shape.        

\begin{table*}
\centering
\caption{\label{tab:3}The predicted numerical results for the $A_{\rm{inv}}$ probe in bins of the $m^{Z \gamma}$ at NNLO QCD at 13 TeV. The predicted results are obtained in different mass regions of the leptonic decay mode as $m^{l^{+}l^{-}}>$ 40 GeV, 100 GeV and 500 GeV. Theoretical uncertainties due to scale variations are included along with the central results.} 
\lineup
%\resizebox{16.0cm}{!}{%
\begin{tabular}{@{}*{11}{c|c|c|c}}
\br
\multicolumn{4}{c}{$A_{Z\gamma}$}\\
\mr
$m^{Z \gamma}$ bin [GeV]  & $m^{l^{+}l^{-}}>$ 40 GeV & $m^{l^{+}l^{-}}>$ 100 GeV & $m^{l^{+}l^{-}}>$ 500 GeV   \\
\mr
95--120         &-0.291	$\pm$-0.0478 	&	0.990	$\pm$0.1943	        &	--                                                           \cr
120--130        &-0.514	$\pm$-0.0746	&	0.945	$\pm$0.1424	        &	--                                                           \cr
130--140 	      &-0.524	$\pm$-0.0594	&	0.855	$\pm$0.0949	        &	--                                  \cr
140--150	      &-0.195	$\pm$-0.0152      &	0.822        $\pm$0.0595   	        &	--                                  \cr
150--160	      &0.101	$\pm$0.0063    	&	0.809	$\pm$0.0541  	        &	--                                 \cr
160--170        &0.230	$\pm$0.0128	&	0.832	$\pm$0.0405             &	--                                                           \cr
170--190        &0.396	$\pm$0.0156	&	0.851	$\pm$0.0299	        &	--                                                           \cr
190--220 	      &0.471	$\pm$0.0219	&	0.865	$\pm$0.0265	        &	--                                  \cr
220--250	      &0.509	$\pm$0.0248      &	0.851	$\pm$0.0331   	        &	--                                  \cr
250--300	      &0.493	$\pm$0.0202  	&	0.823	$\pm$0.0252  	        &	--                                \cr
300--400 	      &0.516	$\pm$0.0277 	&	0.819	 $\pm$0.0352	        &	--                                  \cr
400--500	      &0.519	$\pm$0.0309     &	0.793	$\pm$0.0409   	        &	--                                \cr
500--700	      &0.506	$\pm$0.0314 	&	0.761	$\pm$0.0424  	        &	0.943 	$\pm$0.0489                         \cr
700--2500	      &0.484	$\pm$0.0350  	&	0.716	$\pm$0.0506   	        &	0.803       $\pm$0.0547                          \cr
\br
\end{tabular}%
%}
\end{table*}

\begin{figure}
\center
\includegraphics[width=10.2cm]{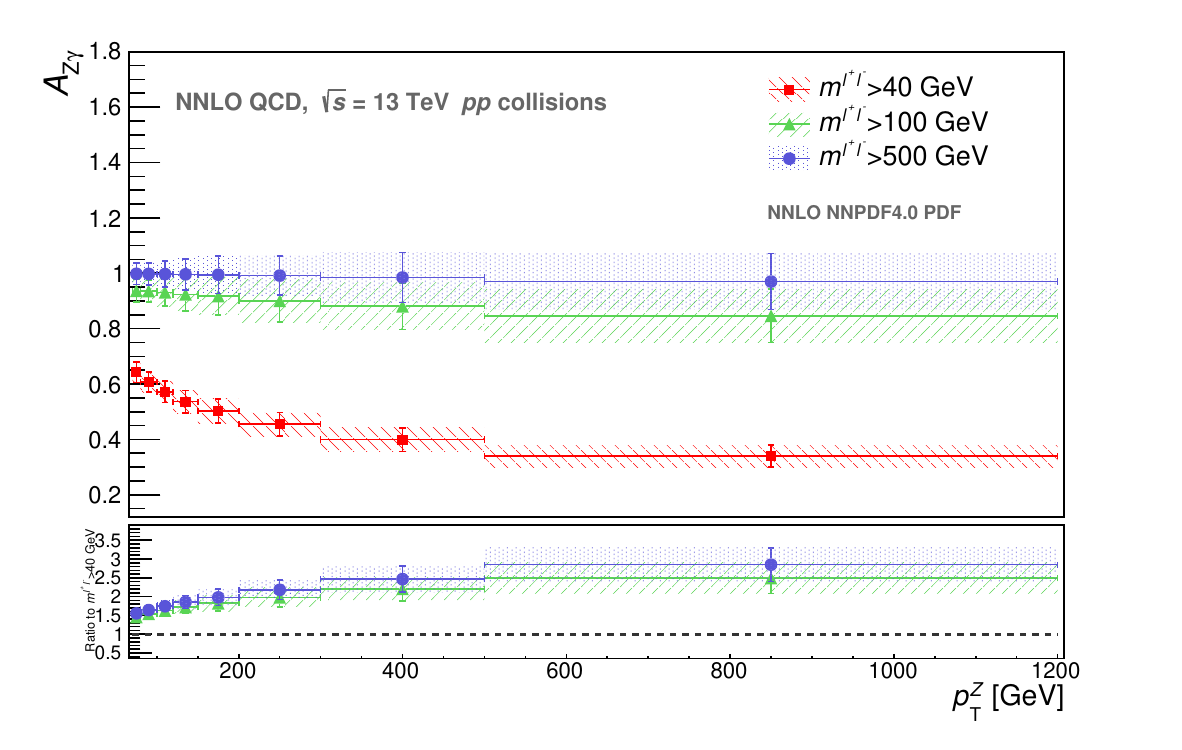}
\includegraphics[width=10.2cm]{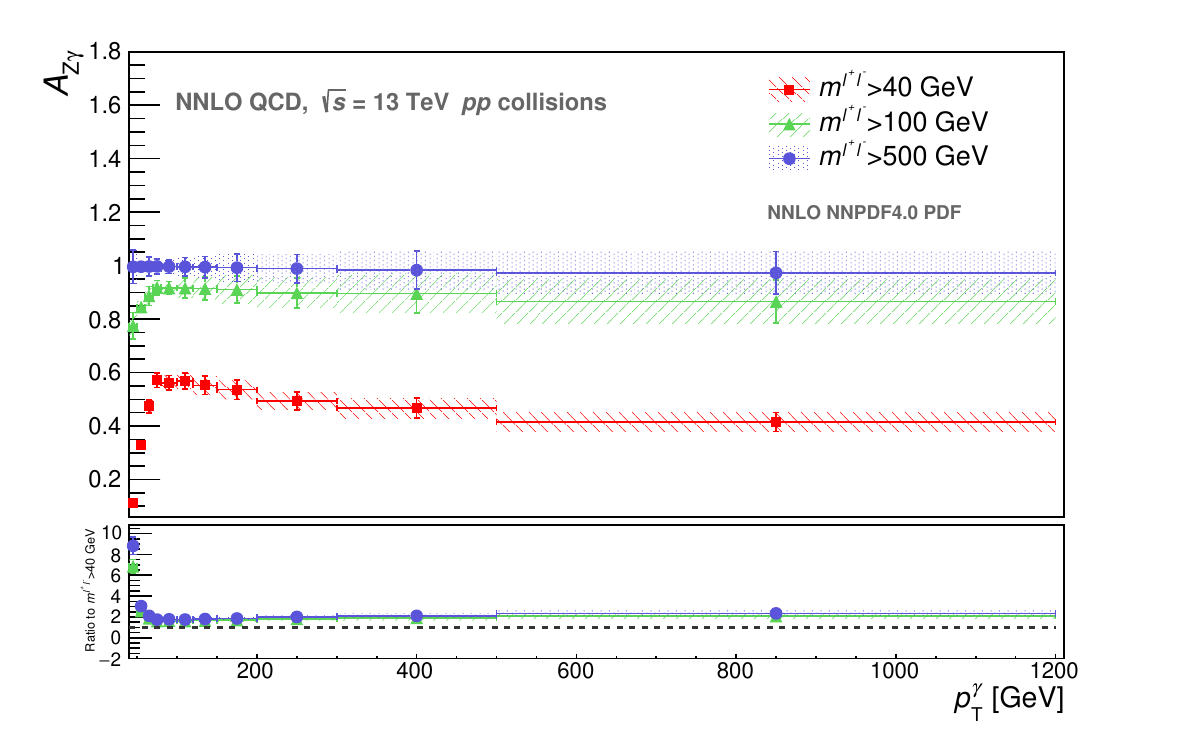}
\caption{The $A_{Z\gamma}$ distributions in the presence of various mass thresholds as $m^{l^{+}l^{-}}>$ 40 GeV, 100 GeV and 500 GeV for the $p_{\rm{T}}^{Z}$ (top) and the $p_{\rm{T}}^{\gamma}$ (bottom), predicted at NNLO QCD at 13 TeV. Theoretical uncertainties due to scale variations are included for the predictions. The ratios of the predictions in higher-mass regions to the predictions with $m^{l^{+}l^{-}}>$ 40 GeV requirement are provided in the lower inset.}
\label{fig:8}      
\end{figure}

\begin{figure}
\center
\includegraphics[width=10.2cm]{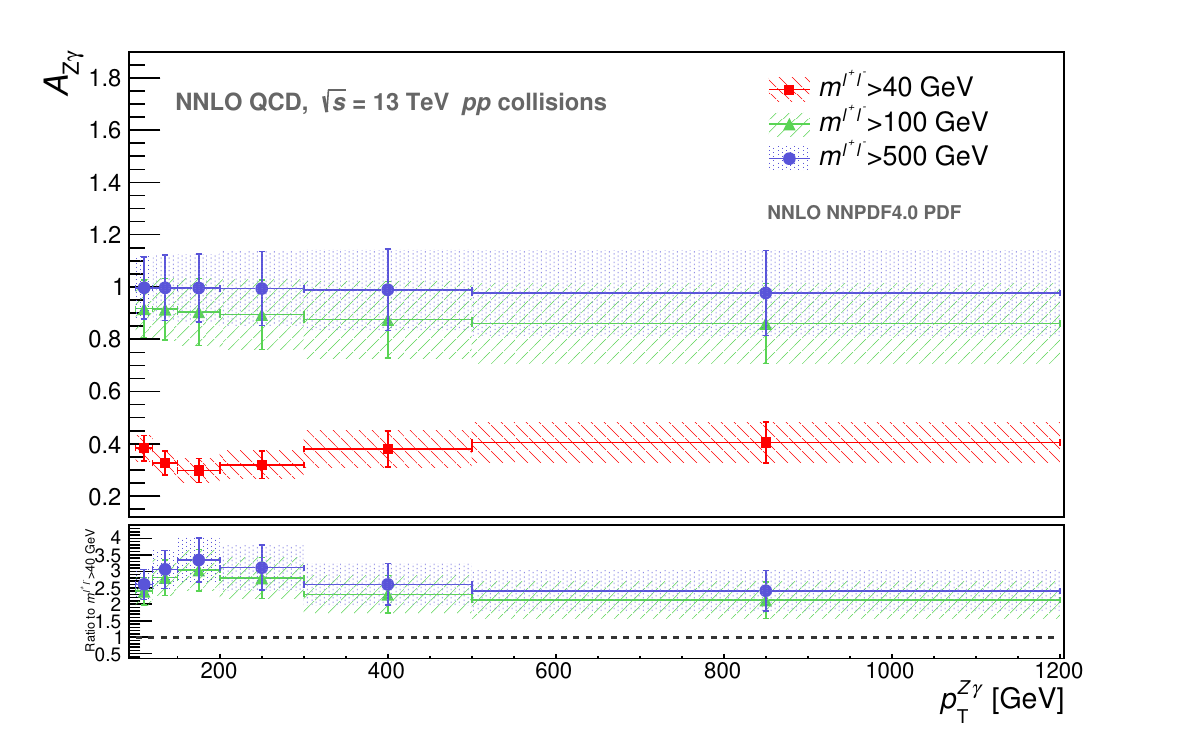}
\includegraphics[width=10.2cm]{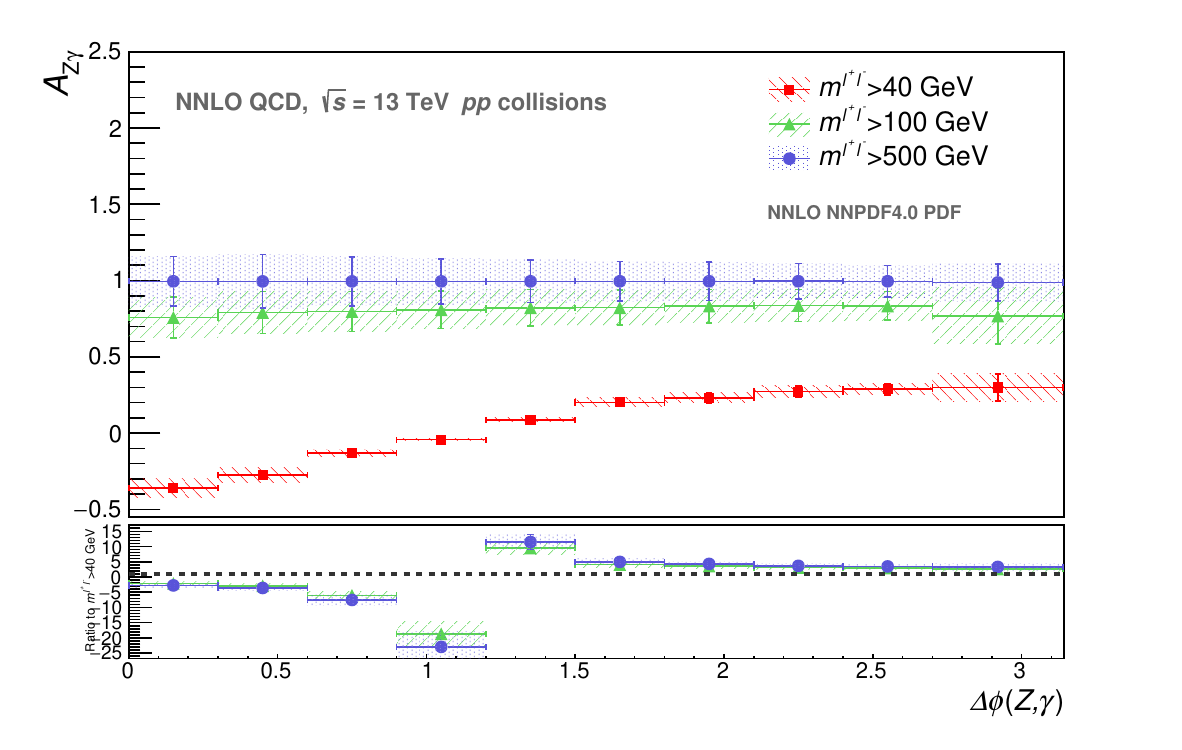}
\caption{The $A_{Z\gamma}$ distributions in the presence of various mass thresholds as $m^{l^{+}l^{-}}>$ 40 GeV, 100 GeV and 500 GeV for the $p_{\rm{T}}^{Z\gamma}$ (top) and the $\Delta \phi (Z,\gamma)$ (bottom), predicted at NNLO QCD at 13 TeV. Theoretical uncertainties due to scale variations are included for the predictions. The ratios of the predictions in higher-mass regions to the predictions with $m^{l^{+}l^{-}}>$ 40 GeV requirement are provided in the lower inset.}
\label{fig:9}      
\end{figure}

Next the key observables $p_{\rm{T}}^{Z}$, $p_{\rm{T}}^{\gamma}$, $p_{\rm{T}}^{Z\gamma}$, and $\Delta \phi (Z,\gamma)$ are subsequently exploited to compare the predicted $A_{Z\gamma}$ distributions in different $m^{l^{+}l^{-}}$ regions. The predicted $A_{Z\gamma}$ distributions are compared for the observables $p_{\rm{T}}^{Z}$ and $p_{\rm{T}}^{\gamma}$ using the requirements $m^{l^{+}l^{-}}>$ 40 GeV, 100 GeV and 500 GeV in Figure~\ref{fig:8}. The distributions are observed to have consistent increasing trend in going from lower $m^{l^{+}l^{-}}$ requirement to a higher one for the entire ranges of the observables. The $A_{Z\gamma}$ has more flat distributions in the high-mass region, specified by the requirements $m^{l^{+}l^{-}}>$ 100 GeV and 500 GeV. Clearly the $A_{Z\gamma}$ exhibit higher sensitivity to high-mass region in higher ranges of these observables, such as when $p_{\rm{T}}^{Z}>$ 200 GeV and $p_{\rm{T}}^{\gamma}>$ 200 GeV. The $A_{Z\gamma}$ reaches almost to a maximum in the highest-mass region $m^{l^{+}l^{-}}>$ 500 GeV in the entire ranges of the both observables. The predicted $A_{Z\gamma}$ distributions in different mass regions are presented for the observables $p_{\rm{T}}^{Z\gamma}$ and $\Delta \phi (Z,\gamma)$ in Figure~\ref{fig:9}. The $A_{Z\gamma}$ distributions increase for the $p_{\rm{T}}^{Z\gamma}$ and $\Delta \phi (Z,\gamma)$ in higher-mass regions and become maximally $\sim$1 in the highest-mass region $m^{l^{+}l^{-}}>$ 500 GeV. In the entire ranges of the $\Delta \phi (Z,\gamma)$, the $A_{Z\gamma}$ values are always positive in higher-mass regions in contrast to the lower ranges of the lowest-mass region $m^{l^{+}l^{-}}>$ 40 GeV. The ratios of the distributions at higher-mass regions to the one at the lowest-mass region $m^{l^{+}l^{-}}>$ 40 GeV are observed to fluctuate in the bin $\Delta \phi (Z,\gamma)=$ 0.6--1.5 owing to very small $A_{Z\gamma}$ values predicted in the lowest-mass region. The $A_{Z\gamma}$ distributions are therefore found to reveal high sensitivities to higher-mass region for all the observables. The $A_{Z\gamma}$ increases and becomes maximum at $\sim$1 towards the highest-mass region $m^{l^{+}l^{-}}>$ 500 GeV, and any observation which will be against these predictions would indicate effects which could only be explained by new-physics models. Consequently, $A_{Z\gamma}$ is assessed to be a good probe for searches for new phenomena including high-mass resonances.    
  
\subsection{The $A_{Z\gamma}$ probe with high-$p_{\rm{T}}^{\nu \bar{\nu}}$}
\label{highptnunu}
The $A_{Z\gamma}$ is predicted at NNLO for phase-space regions in which the $Z\gamma$ process involves the invisible decay into high-$p_{\rm{T}}^{\nu \bar{\nu}}$ neutrino pairs in association with a photon. Impact of high-$p_{\rm{T}}^{\nu \bar{\nu}}$ requirements on the $A_{Z\gamma}$ is potentially interesting owing to presence of large missing transverse energy, as reconstructed by the LHC experiments in reality, resulted from neutrinos which leave experimental apparatus undetected. In addition, the invisible $Z\gamma$ decay is advantageous over the leptonic $Z\gamma$ decay, since the branching fraction for a $Z$-boson decay into a pair of neutrinos is nearly two times higher than for a $Z$-boson decay into a charged-lepton pair, concerning a particular neutrino and lepton flavor. As a consequence, the $A_{Z\gamma}$ is expected to offer higher sensitivity in a phase-space region of the $Z\gamma$ process where it is enhanced by the invisible decay of high-$p_{\rm{T}}^{\nu \bar{\nu}}$ neutrinos. By these token, sensitivity of the $A_{Z\gamma}$ to increasing $p_{\rm{T}}^{\nu \bar{\nu}}$ is assessed by employing the $\emph{Baseline selection--Set II}$ from Table~\ref{tab:2} with different high-$p_{\rm{T}}^{\nu \bar{\nu}}$ thresholds for the invisible decay. The $A_{Z\gamma}$ predictions are compared for the requirements $p_{\rm{T}}^{\nu \bar{\nu}}>$ 60 GeV, 100 GeV, and 200 GeV. QCD scale uncertainties are included in the parton-level predictions. Firstly, the predicted numerical values are acquired in bins of the $p_{\rm{T}}^{Z\gamma}$ as provided in Table~\ref{tab:4}. The $A_{Z\gamma}$ values are observed to follow a decreasing pattern from the lowest-$p_{\rm{T}}^{\nu \bar{\nu}}$ region towards the $p_{\rm{T}}^{\nu \bar{\nu}}>$ 200 GeV region. In the $p_{\rm{T}}^{\nu \bar{\nu}}>$ 100 GeV and 200 GeV regions, the numerical values consistently increase from low-$p_{\rm{T}}^{Z\gamma}$ bin to higher ones up to 1200 GeV. The $A_{Z\gamma}$ thereby constitutes a sensitive probe to different high-$p_{\rm{T}}^{\nu \bar{\nu}}$ regions in describing the transverse momenta of the decay products of the $Z\gamma$ process.    

\begin{table*}
\centering
\caption{\label{tab:4}The predicted numerical results for the $A_{\rm{inv}}$ probe in bins of the $p_{\rm{T}}^{Z\gamma}$ at NNLO QCD at 13 TeV. The predicted results are obtained with different requirements of the invisible decay mode as $p_{\rm{T}}^{\nu \bar{\nu}}>$ 60 GeV, 100 GeV, and 200 GeV. Theoretical uncertainties due to scale variations are included along with the central results.} 
\lineup
%\resizebox{16.0cm}{!}{% 
\begin{tabular}{@{}*{11}{c|c|c|c}}
\br
\multicolumn{4}{c}{$A_{Z\gamma}$}\\
\mr
$p_{\rm{T}}^{Z\gamma}$ bin [GeV]  & $p_{\rm{T}}^{\nu \bar{\nu}}>$ 60 GeV & $p_{\rm{T}}^{\nu \bar{\nu}}>$ 100 GeV & $p_{\rm{T}}^{\nu \bar{\nu}}>$ 200 GeV   \\
\mr
100--120 	      &0.382	$\pm$0.0487	&	0.151	$\pm$0.0168	        &	-0.823	$\pm$-0.0810                                 \cr
120--150	      &0.325	$\pm$0.0461     &	0.174	$\pm$0.0229   	        &	-0.713	$\pm$-0.0768                                 \cr
150--200	      &0.298	$\pm$0.0455    	&	0.220	$\pm$0.0331 	        &	-0.301	$\pm$-0.0376                                 \cr
200--300        &0.319	$\pm$0.0530	&	0.276	$\pm$0.0455             &	0.071	$\pm$0.0109                                  \cr
300--500        &0.380	$\pm$0.0699	&	0.332	$\pm$0.0597	        &	0.255	$\pm$0.0450                                  \cr
500--1200      &0.405	$\pm$0.0781	&	0.384	$\pm$0.0740	        &	0.333	$\pm$0.0631                                  \cr
\br
\end{tabular}%
%}
\end{table*}

\begin{figure}
\center
\includegraphics[width=10.2cm]{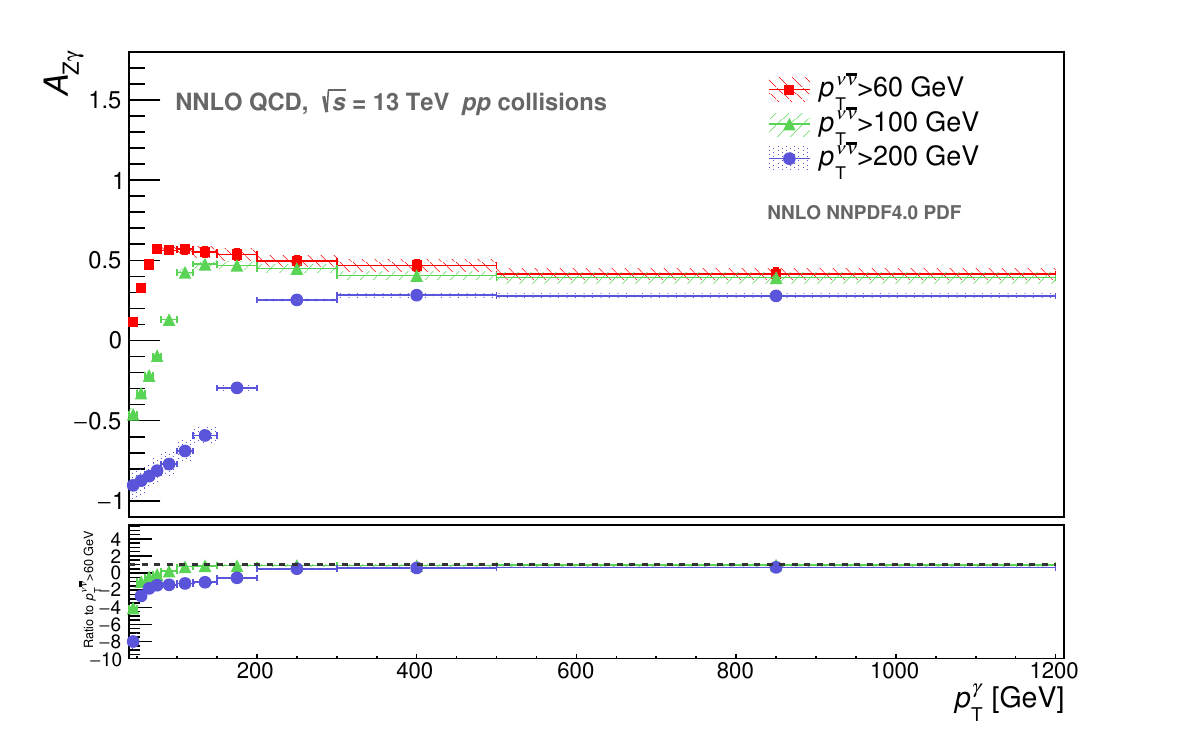}
\includegraphics[width=10.2cm]{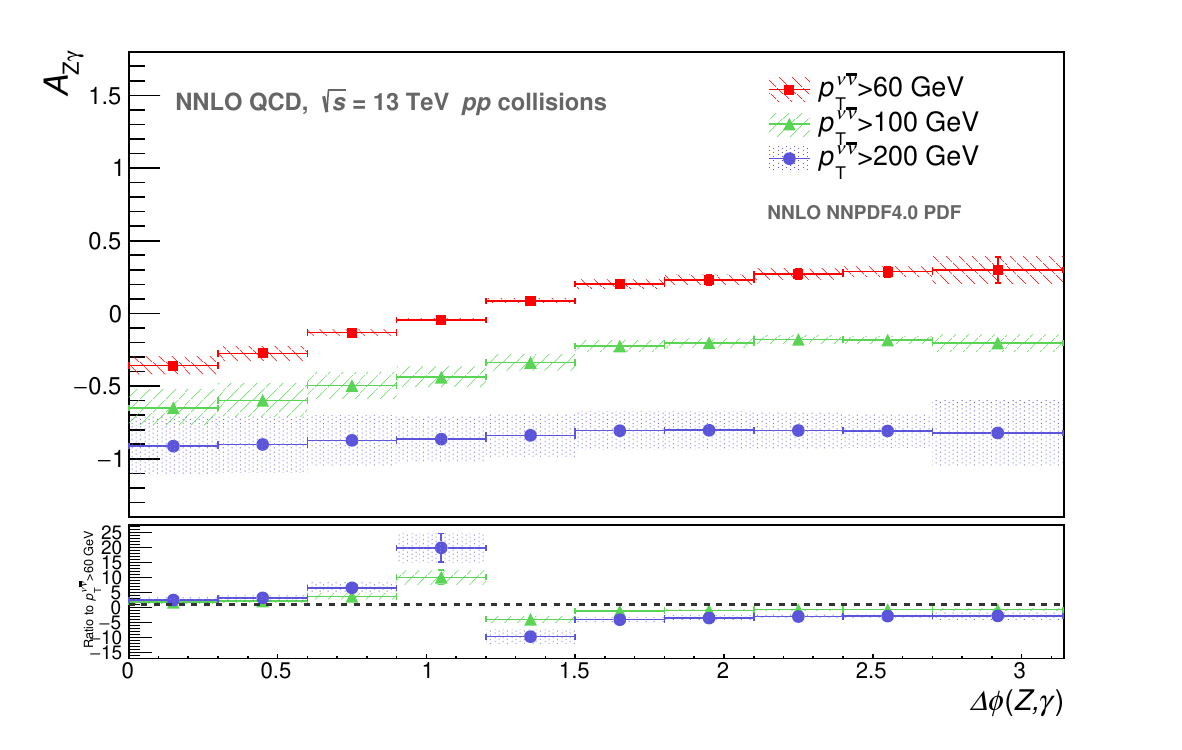}
\caption{The $A_{Z\gamma}$ distributions in the presence of various thresholds for neutrino pair as $p_{\rm{T}}^{\nu \bar{\nu}}>$ 60 GeV, 100 GeV, and 200 GeV for the $p_{\rm{T}}^{\gamma}$ (top) and the $\Delta \phi (Z,\gamma)$ (bottom), predicted at NNLO QCD at 13 TeV. Theoretical uncertainties due to scale variations are included for the predictions. The ratios of the predictions in higher-$p_{\rm{T}}^{\nu \bar{\nu}}$ requirements to the predictions with $p_{\rm{T}}^{\nu \bar{\nu}}>$ 60 GeV requirement are provided in the lower inset.}
\label{fig:10}      
\end{figure}

The predicted distributions in the regions $p_{\rm{T}}^{\nu \bar{\nu}}>$ 60 GeV, 100 GeV, and 200 GeV.are provided for the kinematic $p_{\rm{T}}^{\gamma}$ and the angular $\Delta \phi (Z,\gamma)$ observables at 13 TeV in Figure~\ref{fig:10}. The $A_{Z\gamma}$ as a function of the $p_{\rm{T}}^{\gamma}$ decreases in going from the lowest $p_{\rm{T}}^{\nu \bar{\nu}}$ requirement to the $p_{\rm{T}}^{\nu \bar{\nu}}>$ 200 GeV requirement. In low-$p_{\rm{T}}^{\gamma}$ region such as $p_{\rm{T}}^{\gamma}<$ $\sim$100 GeV ($\sim$200 GeV), the $A_{Z\gamma}$ takes negative values with the higher thresholds $p_{\rm{T}}^{\nu \bar{\nu}}>$ 100 GeV ($p_{\rm{T}}^{\nu \bar{\nu}}>$ 200 GeV). In the region $p_{\rm{T}}^{\gamma}>$ 200 GeV, $A_{Z\gamma}$ distributions turn out to be more flat with the increasing $p_{\rm{T}}^{\nu \bar{\nu}}$ threshold. As a consequence, flatness behavior in the high-$p_{\rm{T}}^{\gamma}$ region renders the $A_{Z\gamma}$ an important probe for observing any small deviation from the SM predictions in indirect searches for new-physics phenomena. No matter how hard neutrino pair of the $Z\gamma$ process escapes a detector, the $A_{Z\gamma}$ is overall predicted to have almost flat distributions in higher ranges of the $p_{\rm{T}}^{\gamma}$ observable. Moreover, the $A_{Z\gamma}$ is inversely impacted by the increasing $p_{\rm{T}}^{\nu \bar{\nu}}$ threshold for the entire region of the $\Delta \phi (Z,\gamma)$. The $A_{Z\gamma}$ distribution as a function of the $\Delta \phi (Z,\gamma)$ becomes minimal and more flat with the $p_{\rm{T}}^{\nu \bar{\nu}}>$ 200 GeV. The $A_{Z\gamma}$ fluctuates numerically in the $\Delta \phi (Z,\gamma)$ range 0.6--1.5 for the distributions with higher-$p_{\rm{T}}^{\nu \bar{\nu}}$ requirements relative to the lowest one. The $A_{Z\gamma}$ is clearly found to be sensitive to the higher-$p_{\rm{T}}^{\nu \bar{\nu}}$ regions throughout the entire ranges of the $\Delta \phi (Z,\gamma)$ observable.           
  
\section{Summary and conclusions}
\label{conclusion}
This paper focuses on a novel approach of exploiting asymmetry between differential cross sections of the leptonic and invisible decays of the off-shell pair-production of a Z boson and a photon in $pp$ collisions, $pp \rightarrow Z\gamma \rightarrow l^{+}l^{-}\gamma$ and $pp \rightarrow Z\gamma \rightarrow \nu \bar{\nu}\gamma$, respectively. Genuine higher-order predictions are presented for the asymmetry of the $Z \gamma$ process $A_{Z\gamma}$ through NNLO accuracy in the context of the QCD perturbation theory. The predicted differential cross sections of the leptonic and invisible $Z \gamma$ decays, based on realistic fiducial phase-space requirements, are validated with relevant data measurements at $pp$-collision energy 13 TeV by the ATLAS experiment at the LHC. Sensitivity of the $A_{Z\gamma}$ is assessed in details at various $pp$-collision energies, in different lepton-pair invariant-mass $m^{l^{+}l^{-}}$ regions of the leptonic decay, and in different neutrino-pair transverse momentum $p_{\rm{T}}^{\nu \bar{\nu}}$ regions of the invisible decay.  

First of all, the predicted differential distributions of the $A_{Z\gamma}$ and of the invisible-to-leptonic cross-section ratio $R_{Z\gamma}$ are presented at various $pp$-collision energies 13 TeV, 14 TeV, and 100 TeV, as functions of a set of key observables including the transverse momenta $p_{\rm{T}}^{Z}$, $p_{\rm{T}}^{\gamma}$, and $p_{\rm{T}}^{Z\gamma}$ and the azimuthal-angle separation $\Delta \phi (Z,\gamma)$ of the $Z \gamma$ decay products. The $A_{Z\gamma}$ appears to be a good alternate of the $R_{Z\gamma}$, where it exhibits slightly varying sensitivities in different regions of phase space of the kinematical observables in comparison with the $R_{Z\gamma}$. The $A_{Z\gamma}$ is found to have notably different sensitivity to the $\Delta \phi (Z,\gamma)$ over the $R_{Z\gamma}$. Sensitivity of the $A_{Z\gamma}$ distributions to increasing collision energies towards 100 TeV is observed to be significant, particularly in kinematic phase-space regions specified by $p_{\rm{T}}^{Z}>$ 200 GeV and $p_{\rm{T}}^{\gamma}>$ 200 GeV. Second of all, the $A_{Z\gamma}$ predictions at NNLO are compared in different $m^{l^{+}l^{-}}$ regions of the leptonic decay of the $Z \gamma$ process at 13 TeV, including $m^{l^{+}l^{-}}>$ 40 GeV, 100 GeV and 500 GeV. The predicted numerical values for the $A_{Z\gamma}$ in bins of the invariant mass of the decay products $m^{Z \gamma}$, are reported to follow consistently an increasing pattern in going from the lowest-mass region $m^{l^{+}l^{-}}>$ 40 GeV towards the highest-mass region $m^{l^{+}l^{-}}>$ 500 GeV. The $A_{Z\gamma}$ distributions are predicted for the key observables $p_{\rm{T}}^{Z}$, $p_{\rm{T}}^{\gamma}$, $p_{\rm{T}}^{Z\gamma}$, and $\Delta \phi (Z,\gamma)$ and compared in different mass regions $m^{l^{+}l^{-}}>$ 40 GeV, 100 GeV and 500 GeV. The $A_{Z\gamma}$ is observed to increase in a consistent manner towards the highest-mass region $m^{l^{+}l^{-}}>$ 500 GeV throughout the entire ranges of all the observables. The $A_{Z\gamma}$ is assessed to reveal high sensitivity in higher $m^{l^{+}l^{-}}$ regions of the $Z \gamma$ process. Sensitivity of the $A_{Z\gamma}$ increases notably in the kinematic region defined by the thresholds $p_{\rm{T}}^{Z}>$ 200 GeV and $p_{\rm{T}}^{\gamma}>$ 200 GeV towards the highest-mass region $m^{l^{+}l^{-}}>$ 500 GeV. And finally, the NNLO predictions for the $A_{Z\gamma}$ are acquired for different requirements $p_{\rm{T}}^{\nu \bar{\nu}}>$ 60 GeV, 100 GeV, and 200 GeV in the invisible decay of the $Z \gamma$ process at 13 TeV. The numerical values of the $A_{Z\gamma}$ in bins of the transverse momentum of the decay products $p_{\rm{T}}^{Z\gamma}$ are predicted to be lowered in going from the lowest-threshold region $p_{\rm{T}}^{\nu \bar{\nu}}>$ 60 GeV to the highest-threshold region $p_{\rm{T}}^{\nu \bar{\nu}}>$ 200 GeV. The $A_{Z\gamma}$ values increase towards higher ranges of the $p_{\rm{T}}^{Z\gamma}$ regardless of the $p_{\rm{T}}^{\nu \bar{\nu}}$ region. The $A_{Z\gamma}$ distributions are predicted for the observables $p_{\rm{T}}^{\gamma}$ and $\Delta \phi (Z,\gamma)$ in the presence of the varying $p_{\rm{T}}^{\nu \bar{\nu}}$ requirements. The $A_{Z\gamma}$ distributions increase with higher $p_{\rm{T}}^{\nu \bar{\nu}}$ requirements, most notably in the low-$p_{\rm{T}}^{\gamma}$ region up to 200 GeV, and exhibit more flat behavior in the ranges $p_{\rm{T}}^{\gamma}>$ 200 GeV in all the $p_{\rm{T}}^{\nu \bar{\nu}}$ regions. The distributions for the $\Delta \phi (Z,\gamma)$ also have increasing trend in going from lowest $p_{\rm{T}}^{\nu \bar{\nu}}$ region to the $p_{\rm{T}}^{\nu \bar{\nu}}>$ 200 GeV. The $A_{Z\gamma}$ is therefore observed to reveal sensitivity to higher $p_{\rm{T}}^{\nu \bar{\nu}}$ regions.                       
   
The $A_{Z\gamma}$ is shown to be a sensitive quantity at higher $pp$-collision energies and in higher $m^{l^{+}l^{-}}$ and $p_{\rm{T}}^{\nu \bar{\nu}}$ regions of the $Z \gamma$ decays, which renders the $A_{Z\gamma}$ an important probe to look for any deviations from the SM predictions as indicators of new-physics effects in indirect searches. New-physics phenomena can be resulted from non-standard interactions of very weakly-coupled light particles, or very heavy particles, or both of them, which have not yet been detected in an experiment. Light particles which couple very weakly to the known particles can manifest themselves in a way similar to the invisible decay of the SM $Z \gamma$ process, while very heavy particles can manifest themselves as high-mass resonances in higher ranges of lepton-pair invariant mass distributions in the leptonic decay of the SM $Z \gamma$ process. These non-SM particles can shelter themselves at high-mass scales which are only accessible at very high $pp$-collision energies such as at 14 TeV or beyond at 100 TeV of the conceived future hadron colliders. In addition, no experimental evidence has been yet reported for anomalous couplings of Z bosons to photons which would be considered as indicators of aforementioned non-SM particles along with the $Z \gamma$ process. In this regards, the $A_{Z\gamma}$ is proposed as an important probe, where it is sensitive to any change in the SM cross sections as a result of very weakly-coupling light or very massive particles at higher collision energies in new-physics scenarios. More specifically, the $A_{Z\gamma}$ can be used for probing very weakly-coupling light particles as predicted in several dark-matter models in the form an enhancement in the invisible decay of the $Z \gamma$ process. The $A_{Z\gamma}$ can also be used for probing any excess in the leptonic decay of the $Z\gamma$ process referring to any massive-particle resonance distributions. Nevertheless, the $A_{Z\gamma}$ is not constrained to be a probe for a few new-physics models, but it can indeed be used in many model-independent searches for new-physics effects. Conclusively, the physics potential of the $Z\gamma$ process can be maximized via the $A_{Z\gamma}$ probe as a sensitive indicator to many new-physics signatures.  

%If you like to have an acknowledgement, uncomment this part
\ack{}
The numerical calculations reported in this paper were fully performed by using High Performance and Grid Computing Center (TRUBA resources) at TUBITAK ULAKBIM.  

\section*{Conflict of interest}
The author declares no conflict of interest throughout the entire paper.

\end{document}